%% file: main.tex
\documentclass[natbib]{svjour3}     
\smartqed  
\usepackage{graphicx}
\usepackage{natbib}
\usepackage[latin1]{inputenc}
\usepackage{txfonts}
\usepackage{color}
\usepackage{calc}
\usepackage{datetime}
\definecolor{cmt}{rgb}{0.0,0.4,0.0}
\definecolor{al}{rgb}{0.6,0.2,0.0}
\definecolor{bl}{rgb}{0.2,0.2,0.6}
\definecolor{sg}{rgb}{0.0,0.0,0.8}
\definecolor{jh}{rgb}{0.0,0.3,0.0}
\definecolor{rc}{rgb}{0.4,0.0,0.0}
\newcommand{\todo}[1]{#1}

\newcommand{\refmark}[1]{#1}
\newcommand{\revmark}[1]{#1}
\newcommand{\fig}[1]{Fig.~\ref{#1}}

\newcommand{\sect}[1]{Sec.~\ref{#1}} 
\newcommand{\secs}[1]{Sections \ref{#1}}

 \journalname{my journal}
\newcommand{\aap}{{Astron. Astrophys.}}
\newcommand{\apj}{{Astrophys. J.}}

\newcommand{\solphys}{{Solar Phys.}}
\newcommand{\pasp}{{Publ. Astron. Soc. Pacific}}
\newcommand{\hinode}{\textit{Hinode}}
\newcommand{\soho}{\textit{SoHO}}
\newcommand{\sdo}{\textit{SDO}}

\newcommand{\kms}{\,km\,s$^{-1}$}

\newcommand{\arcsec}{$^{\prime\prime}$}
\newcommand{\ten}[1]{$\,\times\,10^{#1}$}
\newcommand{\halpha}{H$\alpha$}
\newcommand{\he}{He}
\newcommand{\hei}{He\,\textsc{i}}
\newcommand{\heid}{He\,\textsc{i\,D3}}

\newcommand{\fei}{Fe\,\textsc{i}}

\newcommand{\caii}{Ca\,\textsc{ii}}

\newcommand{\caiihk}{Ca\,\textsc{ii\,h\&k}}

\newcommand{\mgib}{Mg\,\textsc{i}\,B$_2$}

\newcommand{\mgiihk}{Mg\,\textsc{ii}\,h\&k}
\newcommand{\nad}{Na\,\textsc{D}}

\graphicspath{{./}{./figures/}}
\newcommand{\colfig}[3][1.]{\begin{figure}\centering
    \includegraphics[width=#1\linewidth,clip=TRUE]{#2}
    \caption{#3}
    \label{#2}
\end{figure}}

\newcommand{\colfigcr}[3][0.5]{\begin{figure}
  \begin{minipage}{#1\linewidth}
    \centering
    \includegraphics[width=\linewidth,clip=TRUE]{#2}
  \end{minipage}\hfill
  \parbox{0.97\textwidth-#1\textwidth}{
    \caption{#3}
    \label{#2}
     }
\end{figure}}

\begin{document}

\title{Measurements of Photospheric and Chromospheric Magnetic Fields
}


\input{author}

\date{Received: 31-Jul-2015 / Accepted for publication in \textsl{Space Science Reviews}: 23-Oct-2015}

\maketitle

\input{abstract}

\input{introduction}

\input{overview}

\input{groundbased}

\input{spacebased}

\input{chromag}

\input{challenges}

\input{conclusions}
 
\input{acknowledge}


\input{main_bib.bbl}
\end{document}

%% file: author.tex
\author{Andreas Lagg  \and
        Bruce Lites \and
        Jack Harvey \and
        Sanjay Gosain \and
        Rebecca Centeno
}


\institute{A. Lagg \at
              Max Planck Institute for Solar System Research \\
              Justus-von-Liebig-Weg 3\\
              37077 G\"ottingen
              Tel.: +49 551 384979 465\\
              \email{lagg@mps.mpg.de}           
           \and
           B. Lites \at
              High Altitude Observatory, National Center for Atmospheric Research \\
              Tel.: +1-303-497-1517\\
              \email{lites@ucar.edu}
           \and
           J. Harvey \at
              National Solar Observatory \\
              950 N. Cherry Ave. \\
              Tucson, AZ 85719 USA \\
              Tel.: +1-520-318-8337\\
              \email{jharvey@nso.edu}
           \and
           S. Gosain \at
              National Solar Observatory \\
              950 N. Cherry Ave. \\
              Tucson, AZ 85719 USA \\
              Tel.: +1-520-318-8573\\
              \email{sgosain@nso.edu}
           \and
           R. Centeno \at
              High Altitude Observatory,  National Center for Atmospheric Research \\
              Tel.: +1-303-497-1581\\
              \email{rce@ucar.edu}
}

%% file: abstract.tex
\begin{abstract}

The Sun is replete with magnetic fields, with sunspots, pores and plage regions being their most prominent representatives on the solar surface. But even far away from these active regions, magnetic fields are ubiquitous. To a large extent, their importance for the thermodynamics in the solar photosphere is determined by the total magnetic flux. Whereas in low-flux quiet Sun regions, magnetic structures are shuffled around by the motion of granules, the high-flux areas like sunspots or pores effectively suppress convection, leading to a temperature decrease of up to 3000\,K. The importance of magnetic fields to the conditions in higher atmospheric layers, the chromosphere and corona, is indisputable. Magnetic fields in both active and quiet regions are the main coupling agent between the outer layers of the solar atmosphere, and are therefore not only \refmark{involved in} the structuring of these layers, but also for the transport of energy from the solar surface through the corona to the interplanetary space.

Consequently, inference of magnetic fields in the photosphere, and especially in the chromosphere, is crucial to deepen our understanding not only for solar phenomena such as chromospheric and coronal heating, flares or coronal mass ejections, but also for fundamental physical topics like dynamo theory or atomic physics. In this review, we present an overview of significant advances during the last decades in measurement techniques, analysis methods, and the availability of observatories, together with some selected results. We discuss the problems of determining magnetic fields at smallest spatial scales, connected with increasing demands on polarimetric sensitivity and temporal resolution, and highlight some promising future developments for their solution.

\keywords{Sun \and magnetic field \and chromosphere \and photosphere \and measurement \and observations \and spectro-polarimetry}

\end{abstract}

%% file: introduction.tex
\section{Introduction}
\label{intro}

The energy produced by nuclear fusion in the core of our closest star, the Sun, is transported into space almost exclusively by photons emitted from the solar surface. In the absence of solar magnetic fields, this energy flux would likely be nearly constant. It is the solar activity cycle, manifested in the reversal of the solar magnetic field with a 22-year periodicity, which is the main contributor to the variation of the solar energy flux impinging on earth. This variation is only at the order of less than one percent \cite[]{solanki:13a}, but this is sufficient to affect our natural and technical environment in a significant manner. Spectacular eruptions of high energy particles in so-called coronal mass ejections also have the solar magnetic field as their driver. Any reliable space weather forecast, important for example to avoid damage to Earth orbiting satellites, requires the understanding of the energetics in the magnetized solar atmosphere. A robust determination of the solar magnetic field in all atmospheric layers is therefore mandatory.

The transition from the regime dominated by convective gas and plasma motions to the one driven by magnetic forces takes place between the photosphere and the chromosphere. In this layer, the plasma $\beta$ (i.e., the ratio of gas pressure to magnetic pressure) becomes unity. Extremely fast physical processes on size scales at and well below the resolution limit of current solar telescopes, like reconnection or wave dissipation, occur ubiquitously in active regions but also in places of low solar activity. As a logical consequence, solar instrumentation development aims for the determination of the magnetic field at highest spatial and temporal resolution directly above, in and below  the $\beta=1$ layer \cite[][this issue]{kleint:15a}. 

Since in situ measurements of the field using magnetometers in these deep layers of the solar atmosphere are impossible, we have to rely on remote-sensing observations. The information about the magnetic field must be extracted from the radiation emitted from the solar surface and the layers above it. Instruments in the near ultra-violet, the visible regions (where the energy flux from the photosphere and the chromosphere is highest), and the infrared regions of the solar spectrum record the intensity and polarization profiles of Fraunhofer lines. The conditions in the solar atmosphere encoded in these spectral lines are deciphered by the combination of advanced instrumentation and sophisticated analysis techniques.

For more than a century the Zeeman effect has been used to reliably determine the magnetic field in the photosphere \cite[][this issue]{hale:1908,stenflo:15a}. Enhanced instrumental capabilities allowed one to extend the measurements from regions with strong Zeeman signals, like sunspots or pores \cite[][this issue]{rempel:15a}, down to small-scale events and also towards higher atmospheric layers: the chromosphere and the corona \cite[][this issue]{trujillo:15a,wiegelmann:15a}. Especially in these weak field environments, the Hanle effect has gained outstanding importance during the last two decades. In some special spectral lines the advantages of both effects are combined, and allow for continuous measurements from weak, unresolved magnetic fields in the sub-gauss regime up to strong, kilo-gauss field in sunspots.

The unambiguous interpretation of observations using these two effects requires the exact treatment of the physical mechanisms behind them. The radiative transfer equation must be solved, the population of the energy sub-levels in the relevant atoms of the solar atmosphere must be computed, and the effect of radiation and collisions must be considered \cite[][this issue]{delacruzrodriguez:15a}. In addition, the tiny imprints of weak, small scale magnetic fields in the polarization signals of spectral lines require advanced instrumentation with cutting-edge technology in, e.g., optical engineering and detector design.

%% file: overview.tex
\section{Overview of Solar Magnetic Field Measurement}
\label{overview}

Several methods have been used to estimate the properties of the magnetic field in various regions of the solar atmosphere. The assumption that density structures and their dynamics are entrained along or guided by the magnetic field has long been used to infer magnetic properties from observations of the solar corona, prominences and chromospheric fibrils \cite[e.g.][]{bigelow:1889}. Frequencies of oscillation of features thought to be tied to magnetic fields have been used to estimate magnetic field strengths \cite[e.g.][]{hyder:66,ballester:14}. Various properties of solar radio emission are used to deduce magnetic field characteristics in the chromosphere and corona \cite[see review by][]{kundu:90}. In addition to these remote sensing observations, direct measurement of heliospheric magnetic fields has been done for decades, and by 2024 we can expect in situ magnetic field measurements as close as 8.5 solar radii above the solar surface (Solar Probe Plus 2015).
%
%
%

By far the majority of solar magnetic field remote sensing measurements have been made by observations of the Zeeman and Hanle polarization effects on atomic spectral lines. Some reviews of solar spectropolarimetry methods include \cite{stenflo:94a,stenflo:13a} and \cite{keller:02a}. The vastly larger general field of polarimetry measurements and techniques is frequently reviewed. Some recent reviews of special note include \cite{trippe:14}, \cite{snik:14}, and \cite{rodenhuis:14}.

It is useful to consider the key elements of solar spectropolarimetry as a sequence from the source to the inferred measurement: source magnetic field $\rightarrow$ radiative transfer $\rightarrow$ atmosphere $\rightarrow$ telescope $\rightarrow$ calibration $\rightarrow$ \revmark{imaging optics} $\rightarrow$ polarization modulator $\rightarrow$ wavelength selector $\rightarrow$ detector $\rightarrow$ data recording $\rightarrow$ calculation of source polarization $\rightarrow$ inference of source magnetic field properties. Our task is to determine the source magnetic field as functions of 3D spatial position and time from observations of the Stokes vector elements that convey intensity and polarization information as functions of 2D spatial position, spectral wavelength, and time. This is a big challenge. \revmark{In the future, it may be possible to derive more robust 3D information by using simultaneous stereoscopic observations from space.} One could discuss each of the key elements of solar spectropolarimetry in detail but space constraints prevent that. Here we give a simple overview with focused attention on some topics of current interest. 

\subsection{Contemporary Observational and Research Areas}

At present, one may identify four major domains of solar polarimetry research driven by technical tradeoff considerations and scientific interests.

\begin{enumerate}
\item High spatial resolution, with the goal of exploring small-scale magnetoconvection.
\item High polarimetric sensitivity, for studies of \revmark{weak magnetic fields} including the chromosphere and corona.
\item Full solar disk synoptic sequences, for investigations of large-scale magnetic field evolution, the solar dynamo and solar cycle, and for space weather applications.
\item Fast cadence, to better understand solar activity such as flares, CMEs, spicules, filaments, etc.
\end{enumerate}

There is not enough light available to meet all these needs simultaneously using available instrumentation or methods. Tradeoffs are required, and to guide these tradeoffs various system considerations are useful. A spectropolarimetric solar observation can be represented as an 5D hypercube with time ($t$) as one dimension and Stokes vector elements, which represent polarization information ($I,Q,U,V$) as one additional dimension. The remaining three dimensions are two plane-of-the-sky coordinates ($x, y$) and wavelength ($\lambda$). Usually this datacube is drawn at one instant of time with edges $x, y, \lambda$ and cube volume elements containing one of the four Stokes vector elements. Ideally we would like to completely fill such a datacube with a single, brief snapshot in order to obtain observations with the most efficient use of available light. \cite{hagen:13} have very usefully reviewed progress toward snapshot spectral imaging technology and techniques. In practice, current measurements of solar magnetic fields are mostly made using spatially-scanning slit spectrometers or wavelength-scanning narrow band filters. In the former case observing time speed is compromised in order to maintain spectrum line profile integrity at the expense of non-simultaneous spatial sampling. In the latter case observing speed is emphasized but the profile integrity is compromised by the need to observe spectra at different wavelengths and polarizations at different times. The history of magnetic field estimation is full of attempts to mitigate these tradeoffs and their compromises. 

\subsection{Time Constraints}

The duration of an observation is important for two main reasons. First, the Stokes vector must be measured accurately at each datacube element before any observation conditions change. \revmark{Second, the acquisition of a spectropolarimetric data cube must be made sufficiently fast compared to the evolutionary time scale of the magnetic structure observed.} These requirements conflict with both the acquisition of a sufficient number of photons for low-noise Stokes polarimetry and construction of the spatial/spectral images \refmark{quickly} enough to suppress noise from solar feature evolution and ground-based seeing or space-based pointing jitter. Thus, progress in solar magnetic feature measurement is strongly influenced by such tradeoffs, but it is also strongly linked to technology improvements that enhance the efficiency of observations.
In current practice, intensity  measurements in at least four different polarization modulation states are needed to fully define a Stokes vector measurement. A polarization modulator device sequentially produces the different states. Measurement noise arises irreducibly not only from variations in the number of photons detected (shot-noise), but also spuriously from any image intensity or geometry changes during the modulation sequence. A goal is to not add significant spurious noise in excess of the shot noise. For ground-based observations, atmospheric seeing is a major source of noise. \refmark{Measurements of the frequency spectrum of seeing effects on solar images \cite[]{rimmele:00} show large power at low frequencies but also that significant noise is present at frequencies in excess of 100\,Hz.}


\refmark{Practical experience  shows that polarization noise produced by atmospheric seeing may be reduced by rapid acquisition of modulated solar images.} Spatial resolution and site conditions are important factors, but completing a polarization modulation sequence in 10\,ms (from the ground) or a few seconds (from space - depending on the frequency spectrum of pointing jitter) is highly desired for minimizing spurious noise. Ground-based slit spectrograph polarimeters using fast polarization modulation can reach noise levels (uncertainty of $Q, U,$ or $V$ relative to $I$) of order $10^{-5}$ while ground-based filter polarimeters are usually limited to $10^{-3}$ due to an additional time penalty of acquiring many images at different wavelengths and polarization states during which observation conditions change. Adaptive optics and reconstruction techniques for ground-based filter images can reduce the spurious noise caused by seeing. For both ground and space filter observations the evolving pattern of solar intensity and polarization produce polarization noise that depends on the properties of the instrument and the acquisition sequence of the various wavelength and polarization state images. 

\subsection{Solar Polarimetry is Light-Starved}

\revmark{To determine the basic structure of strong magnetic features like sunspots,} noise levels of $10^{-2}$ of the continuum intensity are acceptable. Some high sensitivity work requires noise of the order 10$^{-5}$, or even smaller. Considering just shot noise, 10$^4$ to 10$^{10}$ photons must be collected to reach these noise levels.  Suppose we want to make a photospheric observation at disk center using two spatial pixels per diffraction limit with an overall instrument efficiency of 10\%. \refmark{The temporal modulation typically used for polarization measurements requires us to accept a displacement of no more than 0.05 pixel on the imaging device during the observation, otherwise spurious noise is introduced to the data. Under these conditions and assuming a typical photospheric velocity of 5\kms{},} \fig{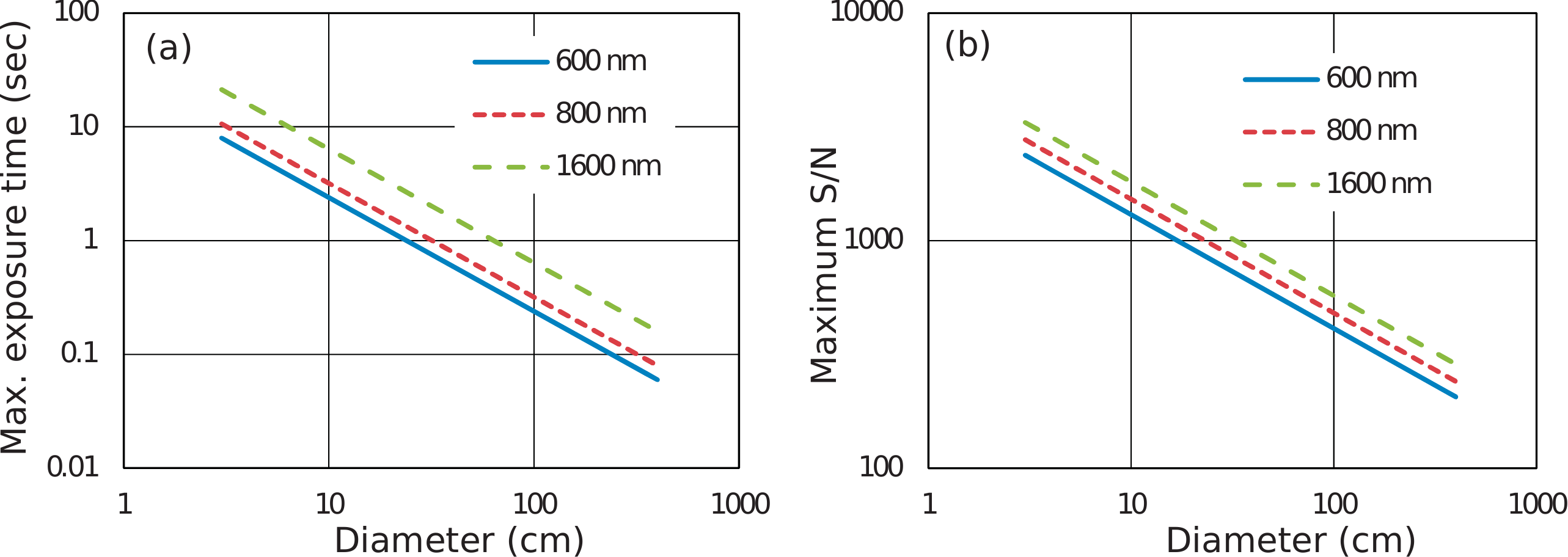}a shows the maximum exposure time allowed at three different wavelengths and a range of telescope apertures. \refmark{At the small aperture size limit shown (GONG) exposure times of about 10 seconds can be used, whilst at the large size limit (DKIST), 60-160\,ms are the longest exposures allowed.}



\colfig{expt_sn.pdf}{Left (a): Maximum exposure time for diffraction-limited disk-center photospheric observation constrained by $<$0.05 pixel motion of target at 5\kms{}. Right (b): Maximum signal to noise ratio per exposure for diffraction-limited photospheric observations at disk center \refmark{resulting from the number of photons per pixel impinging the detector within the maximum exposure time}.}

If we use two spectral samples per spectral resolution element of $\lambda/\Delta \lambda = 90\,000$ and assume an overall system transmission of 10\%, then \fig{expt_sn.pdf}b shows the maximum signal-to-noise ratio (S/N) for Stokes I that can be obtained in the continuum \refmark{computed using the maximum available exposure times}. Note that the S/N ($I/\Delta I$ or $I/\Delta p$) will be lower for $Q,U,V$  ($\Delta p \approx \sqrt{3}/\Delta I$) and in the darker parts of spectral lines. It is obvious that achieving high polarimetric sensitivity in diffraction limited observations is very difficult.

\subsection{Strategies to Improve Efficiency}

The grim outlook shown in \fig{expt_sn.pdf}b has motivated many tradeoffs to be able to secure useful high-resolution, high-precision measurements of solar magnetic fields. Several of these were used in the first solar magnetographs. One may, for example, use larger spatial and spectral samples, combine signals from more than one spectrum line, use sparse sampling of the spectrum, slice the incoming solar image into a format better suited to the spectrometer, and/or select a wavelength compatible with high overall instrumental efficiency. Dual-beam polarimetry was introduced so that all the photons are used all the time. As detector technology advanced, better quantum efficiency and higher read-out speeds were achieved all the time (in addition to greatly reducing crosstalk among polarization states arising from residual image motion \refmark{and blurring}).

There are only a few chromospheric and coronal spectral lines that are useful for magnetography, but many lines forming in the photosphere are available. The photospheric photon flux per spectrum line Doppler width peaks at near-IR wavelengths around 1\,$\mu$m so this should be an optimum wavelength for observations. However, IR spectrum lines tend to be weak compared to those at shorter wavelengths (though this is counteracted by stronger Zeeman splitting at long wavelengths), and spatial resolution for a given telescope aperture is smaller at longer wavelengths. Historically, lines in the visible part of the spectrum have been favored mainly owing to the availability of lower-cost detectors with high quantum efficiency. At present, the optimum wavelength for photospheric magnetic field Zeeman effect measurements ranges from the yellow to the near infrared.

One long-used strategy is to observe only circular polarization (Stokes $V$), from which the line-of-sight component of the magnetic field can be estimated. Provided the magnetic elements are spatially resolved, to first order the strength of the circular polarization signal depends linearly on the source magnetic field strength, while the linear polarization signal (Stokes $Q$ and $U$) has a quadratic dependence, and for modest field strengths it is rather weak.

The availability of detectors with large numbers of pixels that can be read out fairly quickly has stimulated clever techniques to fill the image/polarization/wavelength hypercube in single snapshots \cite[]{hagen:13}. The basic idea is to divide the large space on the detector in ways that allow recording spectral, spatial, and polarization information simultaneously. One older example first used photographically \cite[]{martin:74} is as follows. Instead of using a small format detector in a spectrograph having one long entrance slit, a large format detector allows one to use many entrance slits each producing a spectrum of a different part of the solar image. \fig{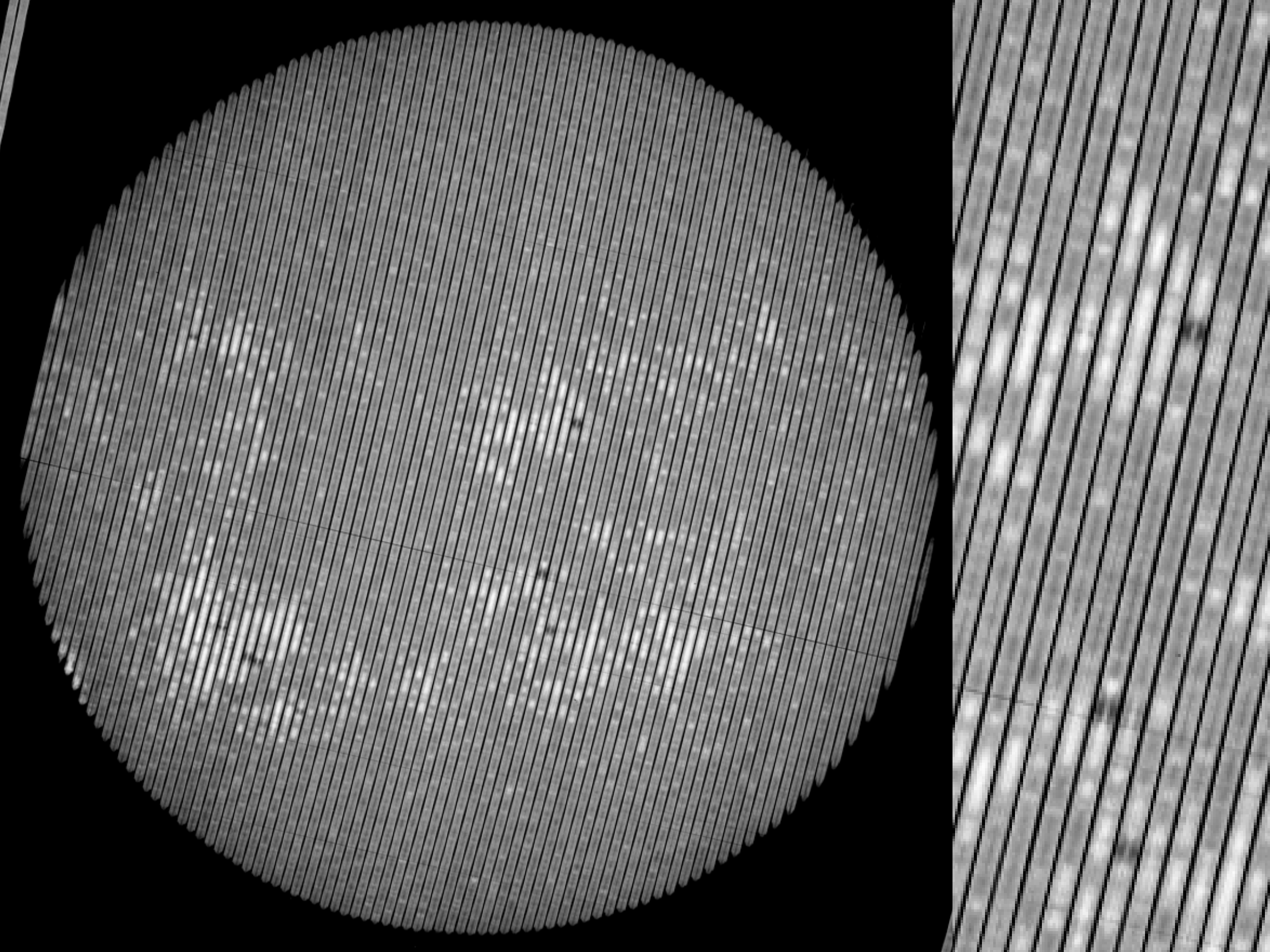} is an early example of what such an observation would look like. Overlapping of the output spectra is mitigated by using a blocking filter with a narrow pass band. So instead of scanning a solar image with one slit, using N slits reduces the required scan time to 1/N, a significant efficiency improvement. There are dozens of other clever ideas to improve efficiency based on large format detectors.  Many of the ideas use fiber optics, microlens, micropolarizer, and microfilter arrays and other modern optical technologies. It is almost inevitable that a new generation of solar magnetographs will utilize some of these efficiency improvements.

\colfig{meudon.pdf}{\refmark{An early example of trading sparse spatial sampling to gain spectral coverage.} Spectroheliogram and zoom showing individual K-line spectra at numerous positions of the entrance slit (Meudon, 1938 August 8).}

Detector performance is key to magnetic field measurement. Fortunately, detector technology continues to improve rapidly. High efficiency sometimes requires conflicting properties. These include: high quantum efficiency over a large range of wavelengths, large pixel count, large dynamic range, excellent linearity, fast readout, no image smear, excellent modulation transfer function, no image lag (signal retention between successive readouts), no etalon fringing, and, if possible, pixel level on-chip demodulation capability. Manufacturers have devised various architectures based on converting photons to charges in a detection layer followed by conversion of the charges to a measurable voltage. In some devices the detection layer is stacked above a layer of readout electronics. In others all the components are essentially in one layer. In some devices (e.g. CCD, most CMOS) the readout process destroys the original charge while in others (e.g. CID, DEPFET, ZIMPOL, sCMOS) the charge is shuffled and not necessarily destroyed immediately by the readout process. To obtain high speed, the sensor area is divided into smaller regions and readout is accomplished simultaneously through multiple channels.

\subsection{State of the Art}

\fig{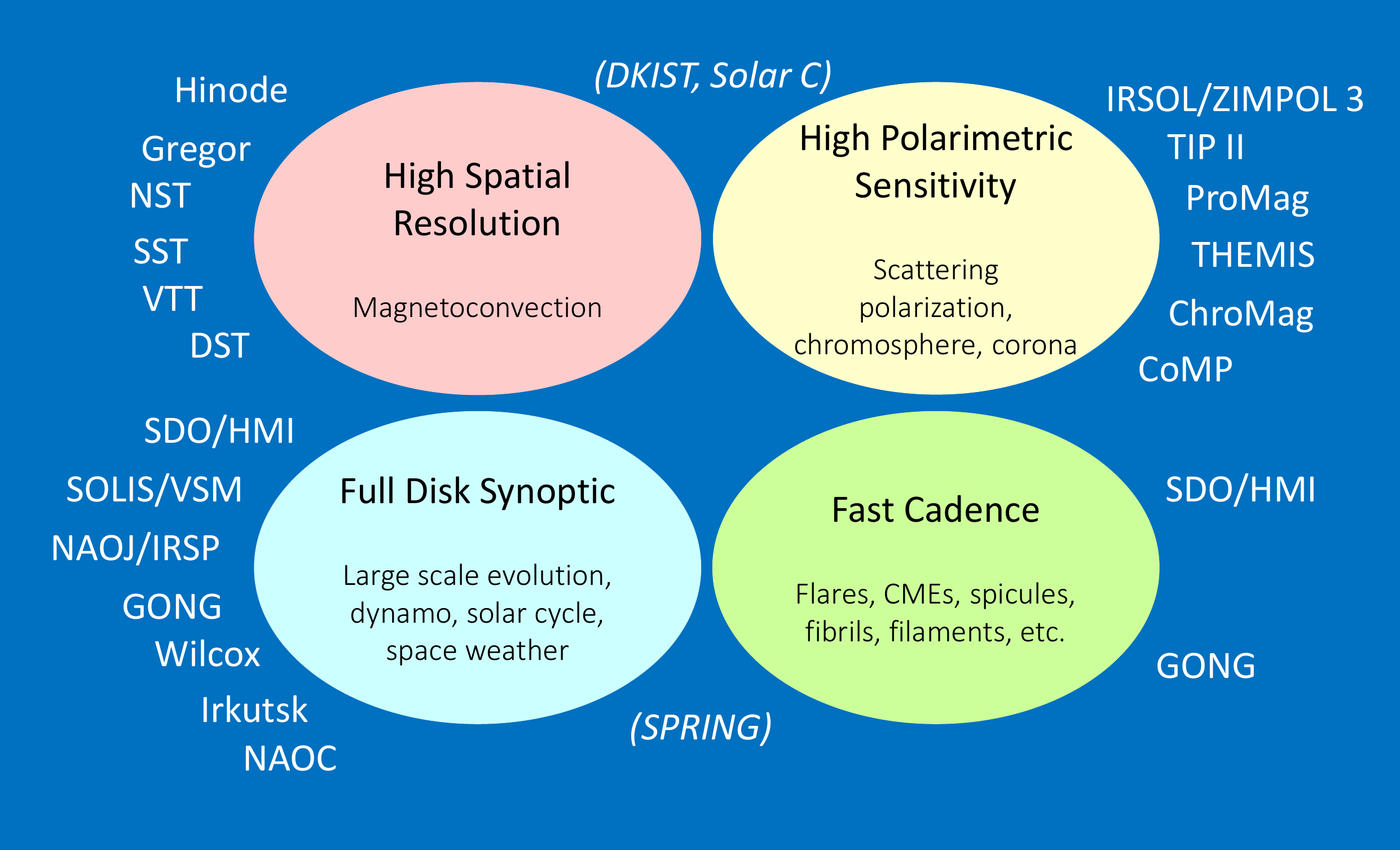} summarizes the major domains of solar spectropolarimetry, and some telescopes/instruments in use (or planned in parentheses) placed close to the domains for which they were designed. This is not a full listing (for example, a few years ago there were more than 20 instruments capable of measuring the chromospheric magnetic field). The figure is intended to suggest the large range of capabilities available for measuring the solar magnetic field limited \revmark{primarily} by financial and personnel resources and the advance of technology.

\colfig{domains.pdf}{Domains of current solar spectropolarimetry and a sampling of telescope/instrument names clustered to the domain for which they are best suited.}

%% file: groundbased.tex
\section{Ground-Based Techniques}
\label{ground}


\refmark{Ground based observations have several advantages over space- or balloon-borne observatories. Their large aperture allows for highest spatial resolution, they can be equipped with complex instrumentation with a diverse selection of observing modes in a multitude of spectral lines, elaborate calibration schemes can be applied, and virtually no limits on data rates do exist. The only impediments to observing from ground in past few decades were the atmospheric seeing effects and the diurnal cycle. The latter has been very successfully mitigated by establishing globally distributed observing networks with identical instrumentation such as GONG, which can be run for extremely long time periods by constant maintenance and upgrades. On the other hand the techniques that are used to overcome seeing effects has seen tremendous progress. In the following we will give examples of the current state-of-the-art in high resolution imaging techniques, narrow band two-dimensional spectrometers, and high precision polarimeters.}

\subsection{Longitudinal Magnetometry: Still a Viable Tool}

\refmark{The term magnetometry has been loosely used in solar physics for the measurement of magnetic fields in the solar atmosphere. However, in reality we can only remotely sense the solar magnetic field by analyzing the photons emitted from the solar atmosphere and applying diagnostic techniques such as Zeeman and Hanle effect to the analyzed photons. Depending on the extent of analysis of the collected photons one can infer either complete vector or just longitudinal component of the magnetic field.} The strength of longitudinal magnetometry is its simplicity in interpreting the measured signal, resulting from the almost linear dependence of the longitudinal Zeeman effect to the line-of-sight component of the magnetic field \cite[][this issue]{stenflo:15a}. Complex, model dependent techniques, like inversions, can often be avoided, and still the fine structure of the solar atmosphere can be retrieved  \cite[]{stenflo:94a}. \cite{babcock:53} used the polarization properties of the Zeeman components of the spectral line to measure the projection of the magnetic field vector along the line-of-sight. The instruments that use this method to map the longitudinal magnetic field of the Sun are popularly known as Babcock type magnetographs. Longitudinal magnetographs have been used extensively for studying the distribution and evolution of magnetic flux on the Sun at a variety of spatial and temporal scales.  Modern longitudinal magnetographs based on two dimensional image sensors exist in various flavors (e.g., SOLIS, \hinode{} SP/NFI, GONG, MDI), and are still very popular for variety of studies such as quiet sun magnetism, flux emergence, active region evolution, magnetic helicity flux estimation, flare related changes, polar field strengths, coronal field extrapolations and so forth.

\subsection{The Age of Vector Magnetic Field Measurements}

In contrast to longitudinal measurements the transverse fields are more challenging to measure and, even more, to interpret. \refmark{The Zeeman effect induced linear polarization signal is proportional to the square of the transverse field strength} and is therefore weaker compared to the circular polarization signal which is proportional to the longitudinal field strength. This different response of the measured signals to the orientation of the magnetic field results often in ambiguous interpretations. For example, the distribution of the magnetic field  orientation in weak field regions is still under debate as to whether the weaker fields are predominantly horizontal or vertical \cite[]{borrero:13}. In active regions, especially near the polarity inversion line, the transverse component of the field becomes strong, often accompanied with a highly non-potential or sheared magnetic field topology. The presence of free or excess magnetic energy above the potential field energy is a necessary condition for the occurrence of energetic phenomena such as flares and coronal mass ejections (CMEs). It is therefore necessary to measure the full magnetic field vector in solar active regions in order to estimate the non-potentiality of the magnetic field and its evolution leading to flare and or CME.
Further, it is now well known that solar active regions show a weak hemispheric bias in their magnetic field twist \cite[]{pevtsov:94}. The twist of the magnetic field can be derived by using various proxies such as force-free parameter $\alpha$ , current helicity, magnetic helicity, chirality of filament barbs, sense of rotation of chromospheric superpenumbral whirls around sunspots.

\refmark{Most of the above mentioned proxies require the measurement of the full magnetic vector field. The origin of the helicity in active regions and their hemispheric bias is still an open question. Various mechanisms have been proposed to explain it, for example its origin in the solar dynamo itself, or during the rise of buoyant flux tubes via their interaction with turbulent convection, or after the flux emergence due to reconnection near the polarity inversion line. Some authors think that the solar dynamo itself is responsible for the generation of helicity, others believe that the turbulent convection interacting with the buoyant flux tubes as they rise towards the surface is the culprit, and a third group thinks that helicity is generated via reconnection processes when the emerged flux interacts with the ambient magnetic properties of the solar corona.}

\refmark{A systematic monitoring of vector fields on the solar disk is therefore needed to understand the relation between magnetic twist and other observables, as well as to build up statistics. Synoptic observations of the magnetic field vector over the full disk derived from full Stokes polarimetry in the photospheric Fe I 630.1~nm line are routinely done by the SOLIS/VSM instrument. \fig{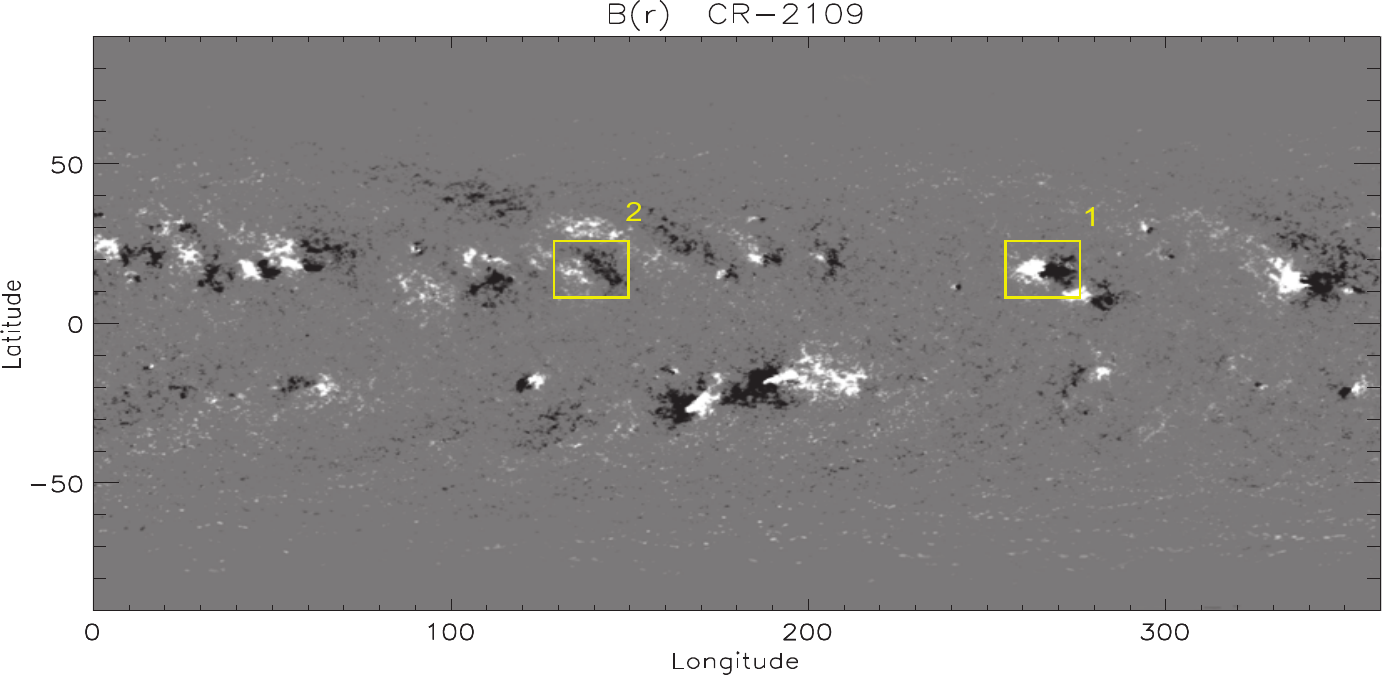} shows the synoptic Carrington map of the photospheric field (radial component B$_r$) generated using the daily fulldisk measurements by SOLIS/VSM \citep{gosain:13}. Such maps are useful, for example, for monitoring the hemispheric helicity trends \citep{gosain:13}, their relation with the kinetic helicity of subsurface flows derived from helioseismology \citep{komm:14,komm:15}, global nonlinear force-free field (NLFFF) extrapolations \citep{tadesse:14a,tadesse:14b} and so on. In addition to synoptic vector field measurements of the full disk it is also desirable to have highly sensitive magnetic field measurements in certain areas of the solar disk such as polar regions and beneath non-active region filaments that also can erupt and be associated with a CME.}

\colfig{sanj_fig1a.pdf}{\refmark{Synoptic Carrington map of the vector magnetic field component B$_r$ synthesized using full-disk SOLIS/VSM vector magnetograms is shown for CR-2109 (the B$_\theta$ and B$_\phi$ components are not shown). The map is scaled between $\pm$100\,G with positive (white) values of B$_r$ pointing upward. The yellow square boxes '1' and '2' show examples of a typical compact active region and a diffuse decaying region, respectively.}}

\subsubsection{Filter-Based or Imaging Instruments\label{filterinstruments}}

As mentioned in \sect{overview} the magnetic field is determined from multi-dimensional spectropolarimetric observations: for every pixel of a 2-dimensional map the spectrum and the temporal evolution should be recorded, resulting in a 4-dimensional data cube for each of the four Stokes components ($I,Q,U,V$).
Since detectors can only record two-dimensional images, a choice of the order of the sampling must be made. Two different approaches are commonly taken, the filter-based and spectrograph-based measurement. Both are eventually leading to same output but are fundamentally different in that the imaging instruments sample two spatial dimensions simultaneously and spectral samples are obtained sequentially, while in the case of spectrograph-based instruments one spatial dimension (along the slit) and the spectral dimension are obtained simultaneously and the second spatial dimension is obtained sequentially by scanning the slit laterally. The fourth dimension of the data cube, time, is achieved by the repetition of these measurements.

\refmark{Examples of \revmark{current} filter type instruments are the Imaging Vector Magnetograph \cite[IVM,][]{mickey:96}, the Interferometric BIdimensional Spectrometer \cite[IBIS,][]{cavallini:06}, the Solar Vector Magnetograph \cite[SVM,][]{gosain:06}, the CRisp Imaging SpectroPolarimeter (CRISP) at the SST \cite[]{scharmer:08}, and the GREGOR Fabry-P\'erot Instrument \cite[GFPI, ][]{denker:10}.} Tunable imaging spectrometers come in various flavors. For example, HMI uses a Michelson Interferometer as a filter, and CRISP, IBIS, GFPI, IVM,  and SVM use tunable air-gap Fabry-P\'erot etalons in single or tandem configuration. The rapid tunability of piezo-mounted air-gap etalons and the high throughput are main advantages of these devices. \revmark{Further, Fabry-P\'erot etalons have the advantage that, unlike for example birefringent filters, they do not contain linear polarizers. This makes dual beam polarimetry possible (see \sect{seeing}) by placing the polarizing beam splitter close to the focal plane of the cameras.} Alternatively, one can also choose solid electro-optic crystal based etalons such as lithium niobate (LiNbO$_3$) crystal wafer based etalons \cite[]{mathew:98,choudhary:02,martinezpillet:11} where the highly polished thin wafer is coated with highly reflective coatings. Here, the wavelength tuning is based on the change of the refractive index \refmark{and the thickness} of the LiNbO$_3$ crystals by applying different voltages. The advantage of these solid etalons compared to normal, air-spaced etalons is their high refractive index ($\approx$2.28), allowing for smaller etalon diameters for a given field-of-view. \refmark{For instance, the largest air-gap etalons (15\,cm aperture) ever made and used in the Improved Solar Observing Optical Network \cite[ISOON,][]{neidig:98} could be replaced by 6.5\,cm aperture LiNb0$_3$ etalons of equivalent throughput. On the other hand their limitations are that they are fragile (specially with larger apertures), have higher absorption losses, and are susceptible to damage if tuned faster than 1000\,Volts/s.}

\refmark{Under the assumption that the acquisition of a spectropolarimetric data-cube is faster than the evolution timescale of the solar structures and that of the atmospheric seeing, filter-based instruments provide a snapshot of the magnetic field over the field-of-view. Thus, such measurements are useful for studies of the temporal evolution of magnetic fields at high cadence, ideal for studies about, e.g., flares or waves. Until the beginning of the new millennium, the magnetic field inference from these observations had some limitations in terms of spectral resolution, sparse wavelength sampling, or spectral purity \cite[]{lites:94}. However, significant progress has been made since then in improving the performance of these instruments in various aspects, for example by combining two or more etalons in tandem configurations to increase the spectral resolution. Instruments like CRISP, IBIS, TESOS and more recently GFPI, therefore now allow to obtain depth dependent, high spatial resolution vector magnetograms. As an example,  \fig{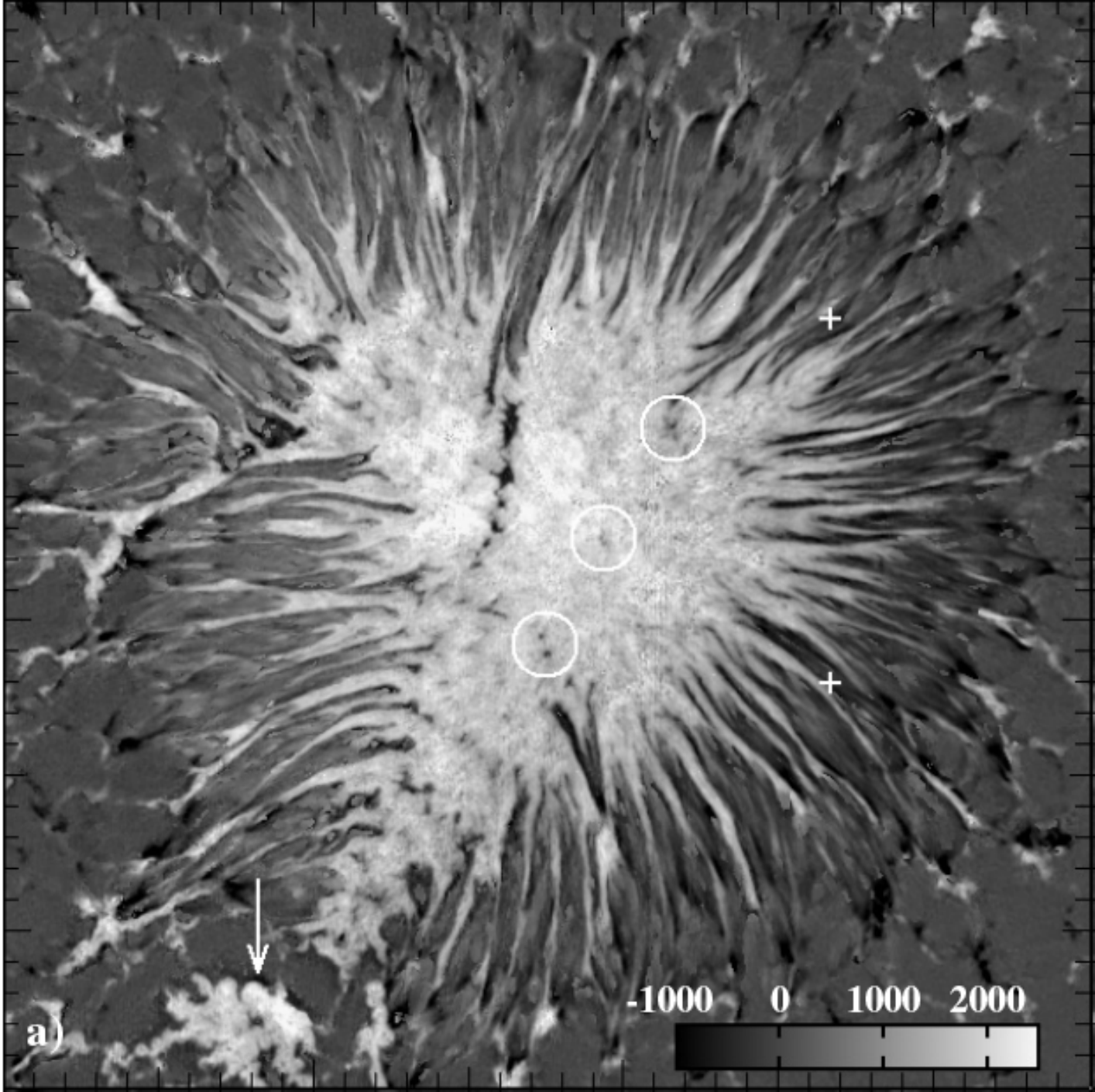} shows the longitudinal field obtained from SST/CRISP observations with twice the angular resolution of Hinode/SOT \cite[reproduced from ][]{scharmer:13}. Such observations are possible by combining several techniques, like adaptive optics, post-facto image reconstruction algorithms and the high spectral resolution and throughput of the CRISP instrument. Due to short acquisition time of the spectropolarimetric dataset from such instruments it is possible to acquire high quality observations even during brief moments of good seeing.}

\colfigcr[0.70]{scharmer_sst.pdf}{\refmark{Longitudinal magnetic field for a 35\arcsec{}$\times$35\arcsec{} large FoV at $\tau_c$=0.95 derived from NICOLE inversions show the presence of opposite polarity fields in most parts of the penumbra. Such opposite polarity fields are absent in higher layers (e.g., at $\tau_c$=0.09, 0.01). The long white arrow points towards the disk center. Figure reproduced from \cite{scharmer:13}, Fig. 4(a).}}

\subsubsection{Stokes Spectrographic Observations}
\label{groundbased:obs}

The spectrograph based instruments sample the spectral dimension simultaneously. Spatial scanning by moving the slit over the field-of-view is needed to build two-dimensional maps. Prominent examples of spectrograph based imaging polarimeters are ASP \citep{lites:96a}, SOLIS/VSM \citep{keller:01}, SPINOR \citep{socasnavarro:06}, TIP \citep{collados:07}, GRIS \citep{collados:13}. In order to expedite the scanning time, a few variants have been developed, such as the multi-slit spectrograph, that uses multiple slits separated far enough so that the usable range of spectra can be recorded without overlap \cite[e.g., FIRS,][]{jaeggli:10}. Here, an order-sorting pre-filter must be used to isolate the wavelength range of interest. Other variants can be grouped into two classes: (1) integral field spectrographs (IFS), where the 2D field-of-view is converted to a one dimensional arrangement feeding a regular slit spectrograph, using either an image slicer \cite[]{bowen:38} or a bundle of optical fibres \cite[e.g. DLNIRSP,][]{elmore:14}, and (2) microlens arrays producing multiple point images out of an extended field-of-view, again feeding a spectrograph with the multiple point sources, oriented such that the dispersion avoids overlapping of the individual spectra (microlens fed spectrograph, M. van Noort, private communication). 
 
\refmark{Simultaneous wavelength information together with typically higher spectral resolution, higher signal-to-noise ratios and finer wavelength sampling are the main benefits of these instruments.} This is of particular advantage when inferring the accurate field strength and orientation in the pixel, including the height stratification of the physical parameters in the line forming region of the solar atmosphere. Sit-and-stare observing modes cater well to the studies of dynamic phenomena, ideally complemented with high temporal resolution slit-jaw imaging. With an IFS the dynamical studies can be extended to a small but 2-dimensional field-of-view. The disadvantage of spectrograph based instruments is that they are not well suited for observations of a large field-of-view  such as a full AR on time-scales of minutes, which is required by some studies such as flare research. 

The ideal solution, therefore, would be to take hybrid approach where the high-quality reflective slit-jaw projection of the spectrograph is used to feed a filter based spectropolarimeter. This allows for information over whole field-of-view with high cadence and limited spectral resolution while the smaller field-of-view sampled by slit (or an IFS) has more detailed spectral information. Real-time processing of the imaging data can help to rapidly re-position the slit to an interesting region, e.g., a flaring kernel or an emerging flux region within the slit-jaw field-of-view.
  
\subsection{Overcoming Seeing\label{seeing}}

One of the biggest challenges for any ground-based measurements is the atmospheric seeing, which changes rapidly within a few tens of milliseconds. The atmospheric seeing can be one of the major sources of spurious polarization signals in sequential measurements \cite[]{lites:87,leka:01}. Dual beam measurements, where the orthogonal polarization states are separated by means of a polarizing beam splitter and recorded simultaneously, are essential for reducing the seeing effects in measurements. Unlike single beam setup, the dual beam setup fully utilizes 100 percent of the photons available for polarimetry. However, different optical paths for the measurement of the two polarization states in a dual beam setup might introduce a systematic differential gain/aberrations error, this puts strict constraints on the optical quality of the two optical paths in such a system. Alternatively, if single beam setup is preferred, a very fast  modulation scheme with a matching fast camera readout must be used, in order to allow a complete modulation cycle before the seeing changes. However, polarization measurements require accumulation of various modulation cycles to build up the desired S/N. This usually results in a loss of spatial resolution, since frames taken under variable seeing conditions must be \revmark{combined} to acquire the desired S/N ratio. In order to preserve or restore the spatial resolution in such observations we then resort to the high angular resolution techniques as described below.

\subsubsection{Image Correction in Real Time}
\label{groundbased:ao}

The optical systems which can compensate for distorting effects of atmospheric seeing in real-time are known as adaptive optics (AO) systems. Such systems typically have a wavefront sensor (WFS), which provides information on the wavefront distortions. This information is used to control in real-time a corrector optics, typically a combination of tip-tilt mirror and a deformable mirror \cite[]{rimmele:11,scharmer:00}. The advantage of such systems is that it (i) enhances the angular resolution in the images, and (ii) allows us to make longer exposures to attain high S/N ratio in polarimetric measurements.  The latter benefit relieves the requirement for extremely fast camera readouts in order to freeze the atmospheric seeing. However, due to the highly variable nature of seeing and due to time constraints in AO systems, only a limited number of higher order \revmark{modes can be corrected}. Thus, the residual seeing effects still remain, introducing seeing-induced cross-talk \cite[]{krishnappa:12}. Higher modulation frequencies help to reduce the seeing-induced crosstalk \cite[]{judge:04,gandorfer:04,ramelli:10,casini:12,krishnappa:12}. 

The seeing is generally expressed with the Fried parameter, $r_0$, which is a measure of the spatial scale over which the wavefront can be considered approximately distortion-free (rms distortion of less than 1 radian). The performance of AO systems  depends heavily upon the basic seeing conditions ($r_0$), the performance being better when $r_0$ is larger. Another property of AO systems is that the quality of real-time corrections is good near the center of isoplanatic patch (typically $\approx$1--2\arcsec{}) in the FoV, the patch which is used for the wave front sensing. A correction over a field-of-view much larger then the isoplanatic patch can be achieved with so called multi-conjugate adaptive optics (MCAO) systems \cite[]{rimmele:10}. First prototypes of MCAO systems are currently being developed \cite[e.g.,][]{schmidt:14}.

Observations in infrared wavelengths, such as the \fei{} 1.56\,$\mu$m lines, are best suited for magnetometry, since the magnetic sensitivity scales linearly with the wavelength. A further benefit of observing in this wavelength region is that the atmospheric seeing varies with wavelength such that $r_0$ is proportional to $\lambda^{6/5}$, i.e., the isoplanatic size is much larger at longer wavelengths.

\subsubsection{Image Reconstruction: Post Processing Techniques}

To further improve the quality of AO corrected data, offline image reconstruction methods, like speckle imaging \cite[]{vonderluehe:93}, phase diversity \revmark{\cite[]{loefdahl:94}}, and multi-object multi-frame blind deconvolution \cite[MOMFBD,][]{vannoort:05} are applied. These techniques are not limited by isoplantism and can be used to improve image quality over a large field-of-view. \revmark{Typically they require separate broadband channel images, recorded at a similar wavelength and strictly simultaneously with the narrowband channel images. The high S/N ratios in these broadband channel images deliver the additional information required for the reconstruction of the narrowband channel images.}

Such reconstruction techniques are essential when narrow band imaging polarimetry is performed as the seeing introduces spatial to spectral crosstalk, and as a consequence spurious polarization signals. Reconstruction methods achieve a good alignment between the pixels of sequentially taken, narrowband images \cite[]{vannoort:05}, which helps avoid spurious polarization signals. The reconstruction techniques are, however, prone to errors or lead to spurious features when the noise in the data is not properly accounted for, and therefore require accurate calibration of parameters. For further details on speckle polarimetry reader is referred to \cite{keller:92} and references therein, and for MOMFBD techniques we refer to \cite{vannoort:05} and \cite{lofdahl:94}.

 
\subsection{Ground-Based Observing Networks}

A network of identical instrumentation distributed geographically around the globe in order to achieve continuous observations of the Sun has been a very successful concept. The long-term operations of networks, such as Global Oscillations Network Group \cite[GONG,][]{harvey:96} or the Birmingham Solar Observing Network \cite[BiSON,][]{chaplin:96}, have led to a continuous data series of uniform quality. The benefit of running such networks are (i) redundancy of sites (e.g., if one site fails there is still flow of data from other nodes), (ii) a possibility of cross calibration and merging of data, (iii) easy upgradeability and maintenance of instruments. Earlier networks were designed for helioseismology studies, which require long, uninterrupted time series of solar oscillation measurements to achieve high frequency resolution. Nowadays, emphasis is put on long term global solar magnetic field measurements for solar cycle studies and continuous monitoring of vector magnetic fields in solar active regions for space weather prediction research.  A proposal to set up a new network of identical instruments designed for measuring fulldisk multi-height velocity and magnetic field vector measurements along with high resolution context imaging in different wavelength bands, such as white light, G-band, H-alpha and Calcium K, is currently funded by European Research Council (ERC). This new network is called  Solar Physics Research Integrated Network Group (SPRING) \cite[]{hill:13}.

\subsection{Examples of the State of the Art and Current Status}
\label{groundbased:status}

The magnetic field observations got a big boost in the last decade owing to the space based \hinode{}/SP and \sdo{}/HMI instruments.  The current next-big-thing in solar magnetic field measurements is expected from the 4-meter aperture Daniel K. Inouye Solar Telescope \cite[DKIST,][]{rimmele:08,tritschler:15}, currently being built on Maui/Hawaii. Equipped with adaptive optics and very sensitive infrared detectors, DKIST will provide magnetic field measurements with unprecedented sensitivity and spatial resolution. However, since the number of photons per diffraction limit resolution elements remains independent of the telescope aperture size, the telescope will be very sensitive when used as a so-called ``photon bucket'', i.e., operated at a resolution coarser than its diffraction limit. Another series of large aperture solar telescopes have become fully operational in the last decade: the 1.6\,m New Solar Telescope \cite[NST,][]{goode:10} at Big Bear, the 1.5\,m GREGOR telescope on Tenerife \cite[]{schmidt:12}, and the 1\,m \refmark{SST} on La Palma \cite[]{scharmer:03}.

All of these telescopes are equipped with fully operational AO systems, while the development of MCAO systems is being carried out in parallel. In the near future, these MCAO systems will allow for magnetic field measurements at a higher spatial resolution over a much larger field-of-view than conventional AO systems deliver \cite[]{rimmele:10}. Also, recent demonstration of ground-layer solar AO by \cite{ren:15} looks promising.

Specialized detectors designed for high speed polarimetry are capable of reaching a polarization sensitivity of the order of 1\ten{-5} of continuum intensity. Examples of such detectors are as below:
\begin{itemize}
\item Charge shuffling cameras, e.g., the Zurich Imaging POLarimeter (ZIMPOL), contain hidden buffers for caching the charges on the chip thus allowing to buildup sufficient signal before the actual readout is performed \citep{povel:94}. Further, since the same pixel is exposed for different modulation states, no differential gain variations are present. The latest version of the ZIMPOL system employs sensors with improved electronics and high sensitivity. New designs/versions of ZIMPOL based on CMOS detectors are also being planned \cite[]{ramelli:10}.
\item The Fast Solar Polarimeter \cite[FSP,][]{feller:14} uses pn-CCD cameras with extremely low readout noise while delivering high frame rates. A polarization sensitivity of below 1\ten{-4} of the continuum intensity has already been achieved. The development of these cameras for larger formats (1 megapixel) is a current project at the Max Planck Institute for Solar System Research.  
\end{itemize}

%% file: spacebased.tex
\section{Space-based Techniques}
\label{space}

\subsection{Motivation for Measurement of Photospheric Magnetic Fields from Space}
\label{motivation}


\refmark{Owing to the circumstance} that many aspects of photospheric magnetism have remained  largely unresolved, the study of solar magnetism has resulted in an ongoing quest  for observations of higher angular resolution.  A fundamental scale of the solar atmosphere is the density scale height ($h_{\rho}$, of order \revmark{150\,km} in the photosphere).  The scale height is the characteristic length for many of the most important hydrodynamic and radiative processes.  Resolving photospheric magnetic fields on this scale has been a goal of solar physics for several decades, and the cleanest way to arrive to a spatial resolution at $h_{\rho}$ would be observations from a space-based platform.  Unfortunately, the cost of deploying a telescope of the necessary size (diameter of order 1\,m) in space has been prohibitive to realization of this goal.  Parallel efforts from ground-based observatories -- larger telescopes, adaptive optics, advanced image restoration techniques -- are pushing observations closer to this long-sought goal.  As a result, the quest for the highest angular resolution from space might now be relegated to a secondary status relative to the benefits of uninterrupted observations with uniformly \textit{very high} angular  resolution.  Provided that continuity of observations is achieved, many extremely important  science problems of solar physics may be addressed effectively without completely resolving the magnetic fields with resolution equivalent to $h_{\rho}$.

Another motivation for space-based observations is access to the hemisphere of the Sun not visible from Earth (or near-Earth orbit), including detailed study of magnetic fields at the solar poles.  Imaging instrumentation has been flown (e.g., the \textit{STEREO} mission), but so far no observations of photospheric magnetic fields have been carried out from such a vantage point, although one such mission is in development (see \sect{sect:future}).

A less obvious motivation for space observations of solar magnetism is support of coordinated observing campaigns.  Science objectives that require diverse observations from multiple observatories (both space- and ground-based) have a much higher chance of success because the space-based components of the campaign have a predictably high probability of success. 

\subsection{Challenges of Space-Based Magnetic Field Measurements}

The development of any space mission is a costly and lengthy process, often necessitating international collaboration in order to make the program possible within budgets of the individual participating countries. International collaborations increase the complexity of management of a space instrumentation program, but they also broaden the pool of  available scientific and engineering talent.  Many of the missions providing measures of photospheric magnetic fields (as summarized below) result from significant international collaboration.

The technical challenges of space instrumentation are significant in comparison to ground-based observing.  Weight and power are always at a premium, as is physical size.  Space instrumentation must be resilient under extremes of temperature and radiation exposure never encountered on the ground.  The vacuum environment can lead to contamination of optical surfaces, so careful attention must be paid to the outgassing properties of materials for space flight.  Furthermore, complex instrumentation must be ultra-reliable and multiply redundant to reduce the possibility of a single-point failure that would bring an end to the mission.  These challenges of space instrumentation drive the high development cost and result in significant compromises in design.   As a rule, the simpler the design and the fewer the moving parts, the more reliable the instrument becomes.  As a result of simplicity, space instrumentation usually has far less flexibility of operation than comparable ground-based instrumentation.

To date, space instrumentation for measurement of photospheric magnetism has carried out measurements of the Zeeman effect in the visible spectrum only.  There is little motivation  for observing photospheric fields in ultraviolet lines because the ratio of magnetic splitting  to the Doppler width decreases linearly with decreasing wavelength.  Furthermore, several other factors render observations at ultraviolet wavelengths more difficult, from both scientific and technical standpoints.  Conversely, the advantages of observing in the infrared are countered by the cost of deploying proportionately larger telescopes in order to achieve spatial resolution equivalent to that of an instrument operating in the visible.

Solar observations require high angular resolution, so the measurement of magnetic fields over the spatial extent of even a single active region demands very high data rates.  Full Stokes polarimetry at multiple wavelengths is necessary for measurement of the vector magnetic field, hence, at least four times the data volume of intensity-only measurements are required.  All of this data must be transmitted to the ground at a cadence appropriate for monitoring the evolution of the target solar features. In order to confront the data requirements in the face of limited telemetry, the polarization data are usually compressed on-board.  The compression leads to some loss of  information that would not be present in equivalent ground-based measurements.  Even with compression, limitations imposed by telemetry remain a limitation to the quantity of data that may be acquired during any given time period.

\subsection{Past and Ongoing Space Missions}

To date there have been only three successful space missions capable of measurements of the photospheric magnetic field.  Two of those, \soho{} and \sdo{}, carried synoptic instruments with full-disk capability intended for precision Doppler velocity measurements for helioseismology in addition to sensing the magnetic field.  The third mission, \hinode{}, has instrumentation optimized for high-resolution study of magnetic fields as a primary objective. These three missions are described in more detail below, and an example of one significant scientific accomplishment is provided for each.

\subsubsection{Solar and Heliospheric Observatory/Michelson Doppler Imager (\soho{}/MDI)}

The \soho{}/MDI instrument \citep{scherrer:95} was primarily intended for helioseismology, but fortunately  the capability for measurement of circular polarization, and hence the line-of-sight component of the magnetic apparent flux density ${B^L}_{\rm app}$, was retained in spite of  pressure to ``de-scope'' the instrument owing to budgetary considerations.  \soho{}/MDI operated from December 1995 to April 2011 and therefore provided coverage of the magnetic field spanning more than one complete solar cycle.

\colfig{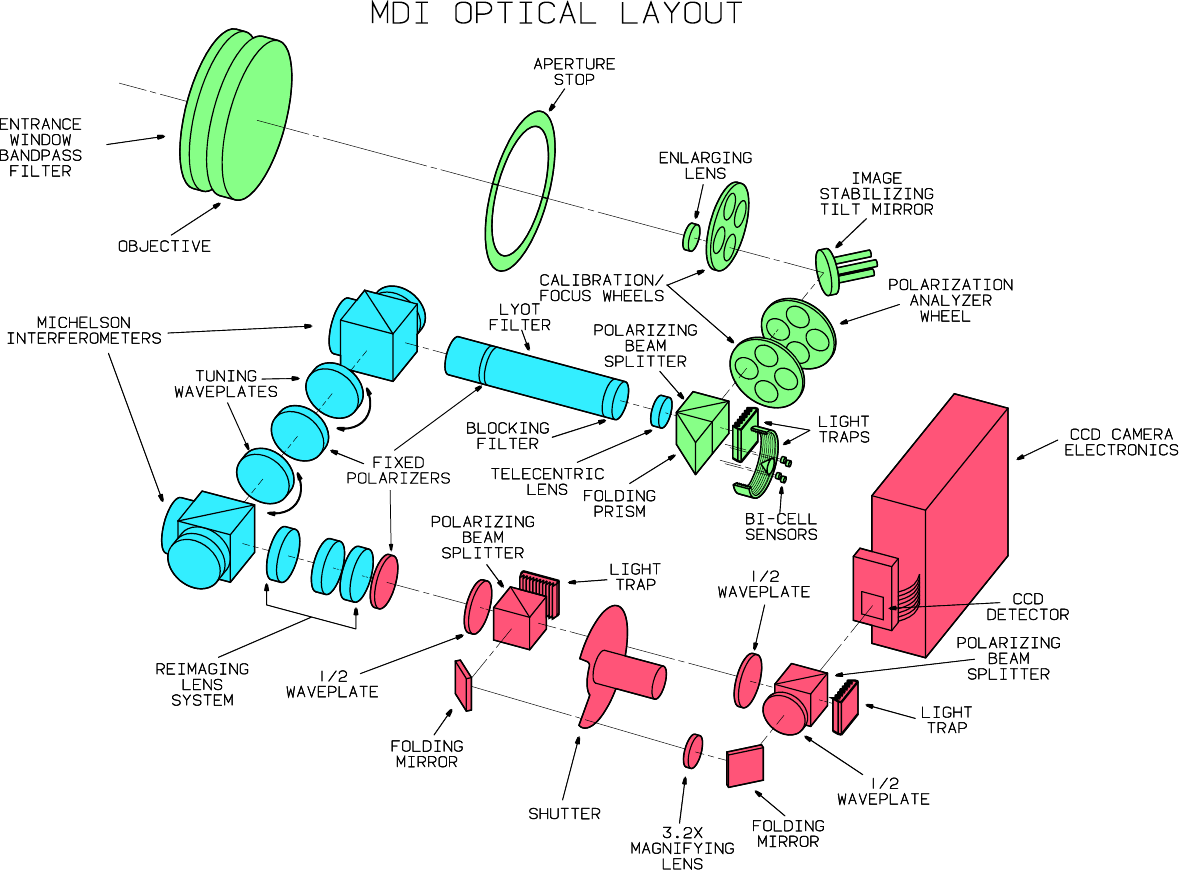}{The optical layout of the \soho{}/MDI instrument is shown.  The wavelength isolation was accomplished by a pair of Michelson interferometers preceded by a Lyot filter.   Circular polarization analysis was accomplished by alternately rotating quarter-wave retarders into the beam with the polarization analyzer wheel.  Diagram taken from \citet{scherrer:95}.}

The optical layout of  \soho{}/MDI is shown in \fig{mdioptical.pdf}. The heart of the instrument, the Michelson interferometers, had heritage from the Fourier Tachometer \citep{brown:81}, a ground-based instrument for helioseismology.  The rotatable waveplates between the interferometers allowed the device to tune over the ~0.5\,\AA{} bandpass of the Lyot filter centered on the Ni\,{\sc i} line at 6768\,\AA{}.  The full-width half-maximum (FWHM) of the combined system was listed at 94\,m\AA{}{}, and in typical operation \soho{}/MDI sampled the line at five wavelengths separated by 75\,m\AA{}.  

The emphasis on helioseismology led to several aspects of this instrument that were not optimal for precision measures of photospheric magnetic fields: performing the polarization analysis in by alternately inserting separate  quarter-wave plates, the use of a spectrum line with less than optimal magnetic sensitivity, a spectral resolution a factor of 2--3 times greater than the typical Doppler width of the Ni\,{\sc i} line in the photosphere, and limited angular resolution (4 arcsec in full-disk mode, 1.2\arcsec{} in high resolution mode).   That being said, owing to its station at the first Earth-Sun Lagrangian point L1, \soho{}/MDI was the first instrument to provide an uninterrupted sequence of longitudinal magnetograms of the Sun  at a regular cadence and uniform quality.  This instrument facilitated the study of the evolution of active regions in a way never before realized.

The magnetogram data from \soho{}/MDI were (and are still) being used extensively to understand solar phenomena, as witnessed by the more than 800 abstracts citing the MDI instrument between 1996 and 2015.  The numerous applications of the data from this instrument facilitated such diverse studies as extrapolation of photospheric fields to compare with structures in the chromosphere and corona, studies of active region evolution including measures of injection of helicity into the upper solar atmosphere, polar fields, flare-induced magnetic field changes, the decay of active regions, and many more. 

The science made possible uniquely by this space experiment is illustrated by the statistical study of the orientation axis of bipolar regions \citep{stenflo:12}.  They examined over 73\,000 active regions from 1995--2011.  This selection included regions with bipolar strength (or flux) ranging from just above the typical flux of ephemeral active regions to the largest observed regions.  They found that the distribution of tilt angles (Joy's law) with solar latitude obeys the same relationship independently of the strength of the dipole (or net flux) \refmark{of the region. 
This behavior led} the authors to conclude that the tilt of active regions is established at the source of the dynamo giving rise to the active regions, not in the buoyant rise through the convection zone as was previously hypothesized.  Furthermore, the authors showed examples (representing just a few percent of the total number of active regions) that did not obey the Hale hemispheric polarity law.  The authors then conclude that these exceptions ``...rule out the possibility of well-defined, coherent toroidal flux systems as a source of all active regions, even the large ones''.


\subsubsection{Solar Dynamics Observatory/Helioseismic and Magnetic Imager (\sdo{}/HMI)}

The Solar Dynamics Observatory (\sdo{}),  launched in February 2011, carries as one of its  complement of instruments the Helioseismic and Magnetic Imager \cite[HMI,][]{scherrer:12,schou:12}.   This instrument is so similar in design to the \soho{}/MDI that no optical layout of the system is provided herein, but it incorporates a number of features that represent significant upgrades from  \soho{}/MDI: 
\begin{itemize} 
\item A different spectral line is observed: Fe\,{\sc i} 6173\,\AA{}.  Not only is this line isolated within a fairly clean region of the spectrum to enable precision helioseismology, but it is also a normal Zeeman triplet with high magnetic sensitivity.
 \item Polarization analysis is carried out with rotating retarders mounted permanently in the beam. This arrangement allows for a higher precision polarimetry than \soho{}\refmark{/MDI}. Additionally, these waveplates allow for measurement of the full Stokes vector, so the instrument is capable of vector magnetometry.
 \item The angular resolution of the instrument is 1\arcsec{} (0.5\arcsec{} pixels), with the field-of-view covering entire solar disk.
\item Two CCD cameras are used, one is dedicated to helioseismology, the other is dedicated to polarimetry.  The data system operates at a higher cadence than that of \soho{}/MDI.
\end{itemize} 
As the name of the instrument implies, this instrument has a much greater scientific emphasis on magnetometry than the \soho{}/MDI did.  Besides continuing the helioseismic function beyond the end of the \soho{}/MDI program, \sdo{}/HMI brings a new dimension in continuous \textit{vector} magnetometry of the full disk.  This instrument makes it possible, for the first time, to perform synoptic studies of the vector magnetic field of every active  region on the disk.  The work of \citet{liu:14} serves as one example of a study involving the \sdo{}/HMI vector magnetogram data of many active regions.  In that work they sought to validate, with data of higher resolution and uniform quality, the rather weak hemispheric preference for  sign of the twist of active region fields as reported by several authors who used ground-based data.   As a measure of twist they adopted a ${B_z}^2$-weighted average of the force-free parameter $\alpha$ (obeying $\nabla \times \bf{B} = \alpha \bf{B}$) over the active region.  \fig{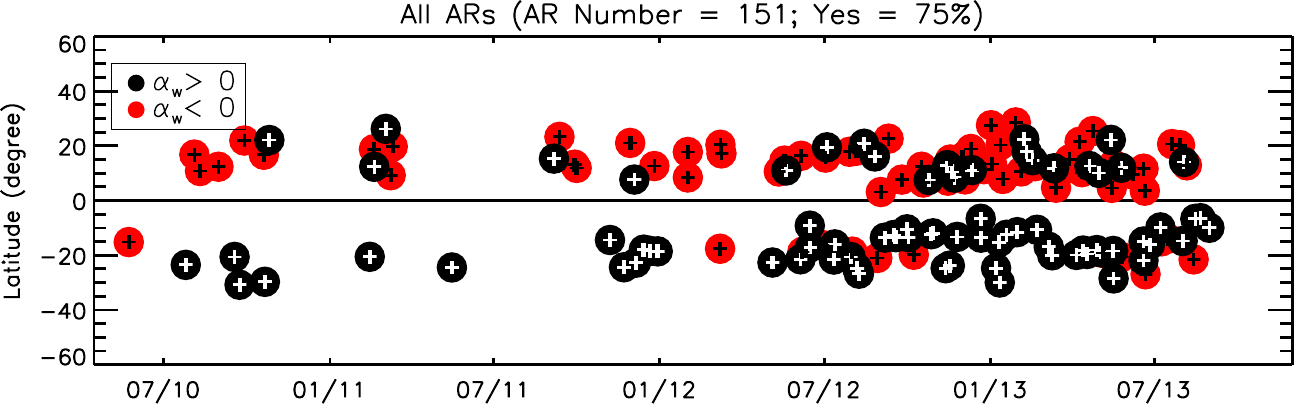} shows the result of their analysis of 151 active regions appearing during the first 3.5 years of the current activity cycle.  This study demonstrates a very clear hemispheric preference for sign of the magnetic helicity.  The high cadence  of \sdo{}/HMI vector magnetic field measurements allowed them to select for study observations of active regions as they traversed the solar central meridian, thereby avoiding any bias that could arise as a result of viewing angle with respect to solar longitude.

\colfig{liuhelicity2014.pdf}{The latitudinal distribution of the sign of the ${B_z}^2$-weighted  average ${\alpha}_w$ of the force-free parameter $\alpha$ is indicated as a function of solar latitude and time during the rise phase of Solar Cycle 24.  The figure results from analysis of vector magnetograms obtained with \soho{}/MDI.  Sign of this proxy of magnetic helicity is indicated by color.  For this distribution of 151 active regions, 75\% show the hemispheric preference for helicity suggested by previous studies of ground-based data. Diagram adapted from \citet{liu:14}.}

Being a very similar cousin to the \soho{}/MDI instrument, the \sdo{}/HMI instrument is optimized for helioseismology rather than precision polarimetry.  As a result, it has some limitations for measurement of the vector magnetic field.  Firstly, the instrument has a rather low duty cycle and therefore it does not use the available photons efficiently. The signal-to-noise ratio of typical polarimetric observations is such that studies of the vector magnetic field in quiet regions are extremely difficult at best. Second, the instrument is a single-beam polarimeter with a slow polarization modulation cycle.  This means that there will be crosstalk among the Stokes parameters arising from uncompensated image motion and evolution of the solar scene.  Third, only one spectral line is sampled, and that with six samples of the line profile.  Each wavelength sample is subject to a filter bandpass of 76\,m\AA{}.  Instruments optimized for Stokes polarimetry sample the complete spectral profile of two or more lines having different sensitivities to the Zeeman effect, and do so with a spectral resolution comparable to the Doppler width of the line in question  (typically 30--40\,m\AA{}~for photospheric lines).  The compromises of the \sdo{}/HMI spectral sampling and resolution result in greatly reduced sensitivity to subtle features of the Stokes profiles that would otherwise assist in precision measures of the magnetic field vector.  They also prevent the accurate extraction of magnetic filling factors in the data reduction process, so the standard \sdo{}/HMI data inversions assume unit filling factor.  For regions outside of sunspots, the extracted field strength is then actually a measure of the average field strength over the resolution element - a measure that is lower than the actual field for the usual case where intense magnetic elements are not resolved spatially. Furthermore, the inferred field inclination will be more transverse to the line-of-sight than the actual field.  Thus one must anticipate significant systematic errors in the inferred magnetic field vector for regions outside of sunspots.  These errors may or may not be significant depending upon the nature of the science question being addressed.  For studies that are highly sensitive to such errors, one should use data from instruments specifically optimized for Stokes  polarimetry, such as the Solar Optical Telescope on the \hinode{} mission described in the following.  Furthermore, limited spectral sampling hinders the extraction of vertical gradients of physical parameters that might otherwise be available to analysis in finely-sampled Stokes profiles, and the limited wavelength range of these samples precludes measurement of strongly Doppler-shifted fields or very strong Zeeman splitting in sunspot umbrae.

\subsubsection{\hinode{} SOT/FG and SOT/SP}

The \hinode{} mission \citep{kosugi:07a} is the first space mission carrying instrumentation specifically optimized for precision Stokes polarimetry.  For this purpose it carries the 50 cm Solar Optical Telescope \citep[SOT:][]{tsuneta:08a} with capability of both spectrographic observations with the Spectro-Polarimeter \citep[SP:][]{lites:13a} and filtergraphic observations with the  Narrowband Filter Instrument (NFI).  The spectrographic observations have the advantages that they provide measures of the spectral Stokes profiles simultaneously at all wavelengths, so that the integrity of the profiles in wavelength is retained, and also that solar events with high Doppler shift do not fall outside of the sampled wavelength range.  These advantages are countered by the non-simultaneity of images of the solar scene that must be built up by stepping the narrow spectrograph slit across the image.  One may have the best of both worlds with simultaneous filtergraphic and spectrographic Stokes polarimetry as implemented in the SOT/SP/NFI instrumentation.

\colfig{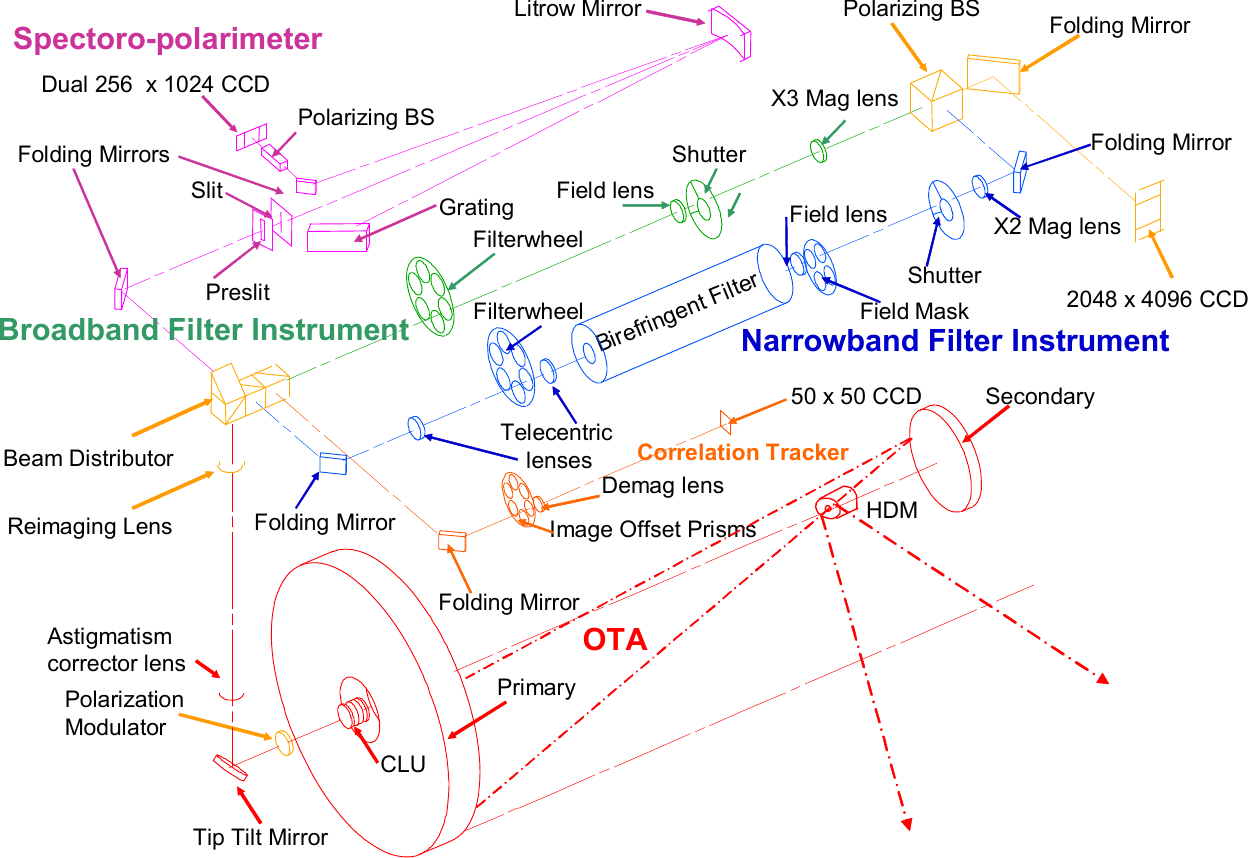}{The optical layout of the \hinode{} SOT identifies the five major optical systems with different colors.  Of particular interest for this paper are the Narrowband Filter Instrument (blue) and the Spectro-Polarimeter (magenta), both of which are specifically designed to  perform precision polarimetry for measurement of solar magnetic fields.  For a detailed description of the Spectro-Polarimeter see  \citet{lites:13a}. Diagram taken from \citet{tsuneta:08a}.}

The optical layout of the SOT Focal Plane Package (FPP) is shown in \fig{sotoptical.pdf}. Polarization modulation is accomplished by a rotating bi-crystalline retarder (quartz and sapphire) located in the symmetric part of the optical system just after the Optical Telescope Assembly (OTA).  After the reflection by the tip-tilt mirror the  beam is divided among four optical paths:  the SP, the Broadband Filter Instrument (BFI),  the NFI, and the Correlation Tracker.  The NFI and BFI share a common focal plane and detector.  The SP and NFI (or BFI) can operate simultaneously because the SP has its own detector.  The polarization modulator rotates at 0.625\,Hz such that the frame transfer detectors sample continuously at 10\,Hz: a sampling speed fast enough to ``freeze'' moving features in the solar atmosphere for most polarimetric measurements.  The SOT/SP operates in dual-beam mode by imaging both orthogonal linear polarizations exiting the  polarizing beam splitter near its focal plane. Owing to continuous integration/readout of the detector, the instrument operates at close to 100\% duty cycle.  Several modes of polarimetry are available to the NFI, including frame-transfer modes using an externally positioned mask of the focal plane.  So both SOT/NFI and SOT/SP are capable of high precision polarimetry at high angular resolution.

\colfig{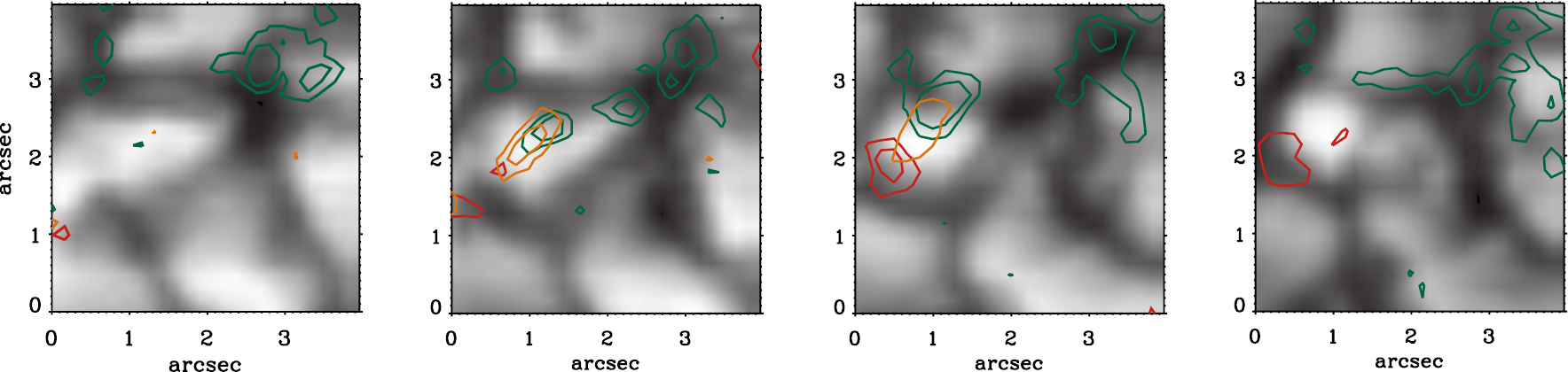}{This sequence of four continuum intensity images derived from repeated,  short maps of the \hinode{}/SOT Spectro-Polarimeter reveal the emergence of a small dipole loop in a granule of the quiet Sun.  Red and green contours represent positive and negative circular polarization (upward- and downward-directed magnetic field), and orange contours indicate linear polarization (horizontal field component).  A loop  structure is seen to form in the second frame with horizontal fields between the two  vertically-oriented fields of opposite polarity.  As the top of the loop rises above the photosphere, only vertical fields remain in the photosphere, each centered over a dark intergranular lane. Diagram taken from \citet{centeno:07}.}

Freedom from the disturbing effects of the Earth's atmosphere allows the \hinode{} polarimeters to continuously achieve angular resolution of the 50 cm SOT.  This has permitted the detailed study of small-scale magnetism in the solar atmosphere.  One area of intense study by \hinode{} researchers has been the magnetism of the quiet solar atmosphere, as illustrated by the example in \fig{centenoemerge.pdf}. Prior to \hinode{}, glimpses from ground-based instrumentation revealed the presence of small-scale horizontal fields in the solar photosphere \citep{lites:96}, and it was suspected that they might correspond to emergence of small magnetic loops into the photosphere.  With the advent of \hinode{} the widespread occurrence of these features was revealed for the first time \citep{lites:08a} owing to the excellent image quality combined with high polarimetric precision.  Using repeated short maps of quiet regions with SOT/SP, \citet{centeno:07} were able to demonstrate that the horizontal fields are indeed the tops of tiny magnetic loops emerging within granules, as shown in \fig{centenoemerge.pdf}.

\refmark{As a consequence of the seeing-free space environment, the point spread function (PSF) does not vary significantly during the time of an observation. This attribute allowed the development of inversion codes that take into account the PSF self-consistently during the minimization process, with the result that maps having diffraction-limited resolution of the SOT may be obtained for the physical parameters in the solar atmosphere. Such techniques are described in \cite{vannoort:12,asensioramos:15} and also in this special issue \cite[]{delacruzrodriguez:15a}.  They have already resulted in numerous publications \cite[e.g.,][]{lagg:14,tiwari:13,buehler:15}.}

The \hinode{} mission was launched in September 2006 and still continues to  operate providing measures of the photospheric magnetic field reliably and on demand.  It is the only such space mission operating now, and comparable or better instrumentation is only on the horizon, being at least 6--10 years away.  The SOT/SP is operating nominally, having experienced only a 25\% drop in throughput over the past nine years on orbit.  The SOT/NFI instrument continues to provide limited  magnetogram data in the wing of the Na D-line only, owing to a progressive drift off-band of the pre-filters to the birefringent filter.  Also, about a year into the mission, \hinode{} suffered a failure of its high-rate X-band telemetry.  Fortunately, the lower rate S-band telemetry continued to function well, and conservative management of the data volume from SOT combined with a greatly increased number of downlink passes has allowed most of the SOT science objectives to continue to be addressed.  This history of \hinode{} telemetry issues is yet another testament to the need for redundancy in space mission hardware.

\subsection{High-Altitude Balloon Missions}

Although they are not strictly space missions, a brief discussion of high-altitude  balloon missions carrying instrumentation to measure the photospheric magnetic field is provided here.  Balloon platforms reproduce some desirable conditions of space: 1) the rarified atmosphere at high altitude nearly completely eliminates the disruptive effect of atmospheric seeing, 2) during the high-latitude mid-summer extended periods of solar observing without night interruption are possible, and 3) some ultraviolet wavelengths become nearly transparent.

The cost of a balloon experiment is much smaller than that of an equivalent experiment in space because of relaxed restrictions on weight, power, contamination, vibration/acoustic tolerance, documentation, and other factors.  Cost issues aside,  development of a balloon mission is still a major undertaking.  Taking the the two prior solar magnetographic balloon missions as typical,  the development time for a major long-duration balloon experiment is comparable to that of a space mission.  Reliability constraints of a balloon mission are similar to  those of a space experiment since high altitude balloon launches are expensive and infrequent.  One usually has the opportunity to recover the flight hardware after the end of a balloon mission, but that hardware must  be able to survive a crash landing if it is to be flown again. In most respects, pointing and image stabilization of a balloon-borne high resolution telescope is more difficult than a space platform because one has to contend with variable winds at float altitude, and variable wind gradients between the balloon itself and its payload drive a pendulum motion.

Solar long-duration balloon experiments may be flown at mid-summer in the polar regions in order to have continuous observing throughout the flight.  Two solar magnetographic experiments have been flown: the Flare Genesis Experiment \citep[FGE,][]{bernasconi:00, bernasconi:01, bernasconi:02}, \refmark{and the successful Sunrise program as described in the following.}

\subsubsection{Sunrise/IMaX}

\refmark{The Sunrise program} \citep{barthol:11} \revmark{is} a high-altitude, long-duration balloon experiment intended to explore solar phenomena at very high angular resolution \cite[see also ][in this issue]{kleint:15a}.  It was a joint effort of Germany, Spain, and the USA.  It consisted of a 1.0\,m telescope carrying the Imaging Magnetograph eXperiment  \cite[IMaX - \fig{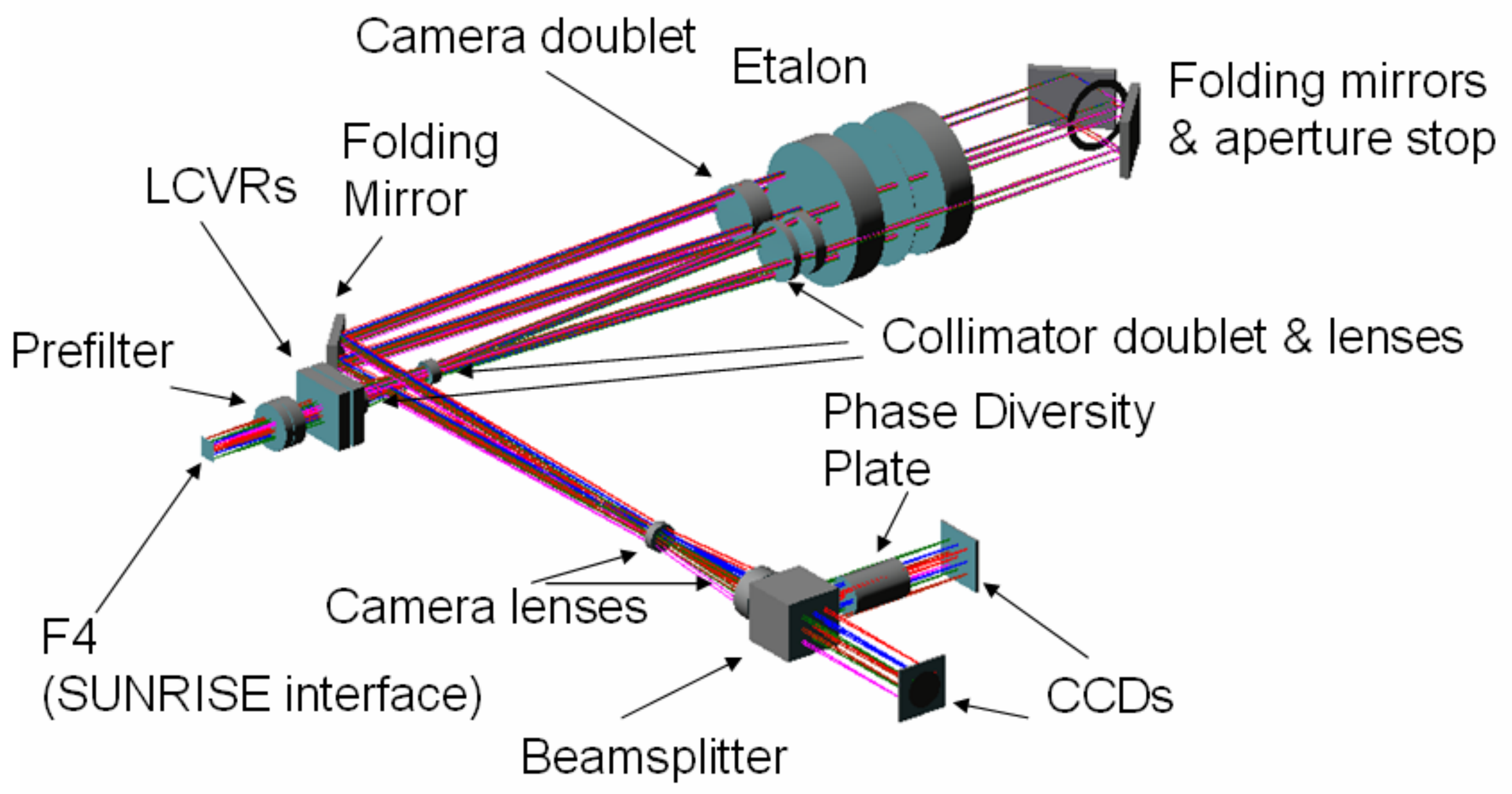},][]{martinezpillet:11} and an ultraviolet imager \cite[Sunrise Filter Imager (SuFI),][]{gandorfer:11}.  The former operated in the highly Zeeman-sensitive Fe{\sc i} line at 5250\,\AA{}.  Also like FGE, IMaX used a pair of liquid crystal variable retarders to accomplish polarization modulation, but as a polarimeter IMaX had many features that qualified it as a quantitative solar polarimeter. IMaX measured combinations of four states that uniquely define the Stokes 4-vector.  Compared to a polarimeter measuring pure states with six measurements, the IMaX scheme has significantly higher polarimetric efficiency \citep{deltoroiniesta:00}.  Furthermore, the instrument had a nearly continuous expose/readout cycle resulting in a duty cycle of near unity.   These  features allowed IMaX to perform polarimetry at a S/N appropriate for quantitative determination of the magnetic field vector.  IMaX also used a tunable lithium niobate Fabry-P\'erot filter in double pass so that it achieved a spectral resolution of 85\,m\AA{} \revmark{at 5250\,\AA{}}. In typical full-Stokes polarimetry mode it sampled five wavelengths throughout the line profile, each with a S/N of about 1000. 

\colfig{imax_optical.pdf}{The optical layout of the IMaX instrument on the Sunrise balloon program illustrates the simplicity of its design.  Having two focal planes, it may operate as either a dual-beam polarimeter or, by applying a slight de-focus to one of the beams, as a phase-diversity imager.  The wavelength isolation is accomplished by a tunable lithium niobate crystalline Fabry-P\'erot etalon in double pass. Diagram taken from \citet{martinezpillet:11}.}

Two flights of Sunrise have been realized.  The first flight in the north polar region lasted six days in June 2009, and produced outstanding results.  High resolution  sequences of data of duration up to 30 minutes were obtained of quiet Sun conditions. The Sunrise/IMaX system achieved close to its design angular resolution of about 0.15\arcsec{}, i.e. a factor of two better than that of \hinode{} SOT/SP.

An example of the science that was made possible by Sunrise/IMaX is illustrated by \fig{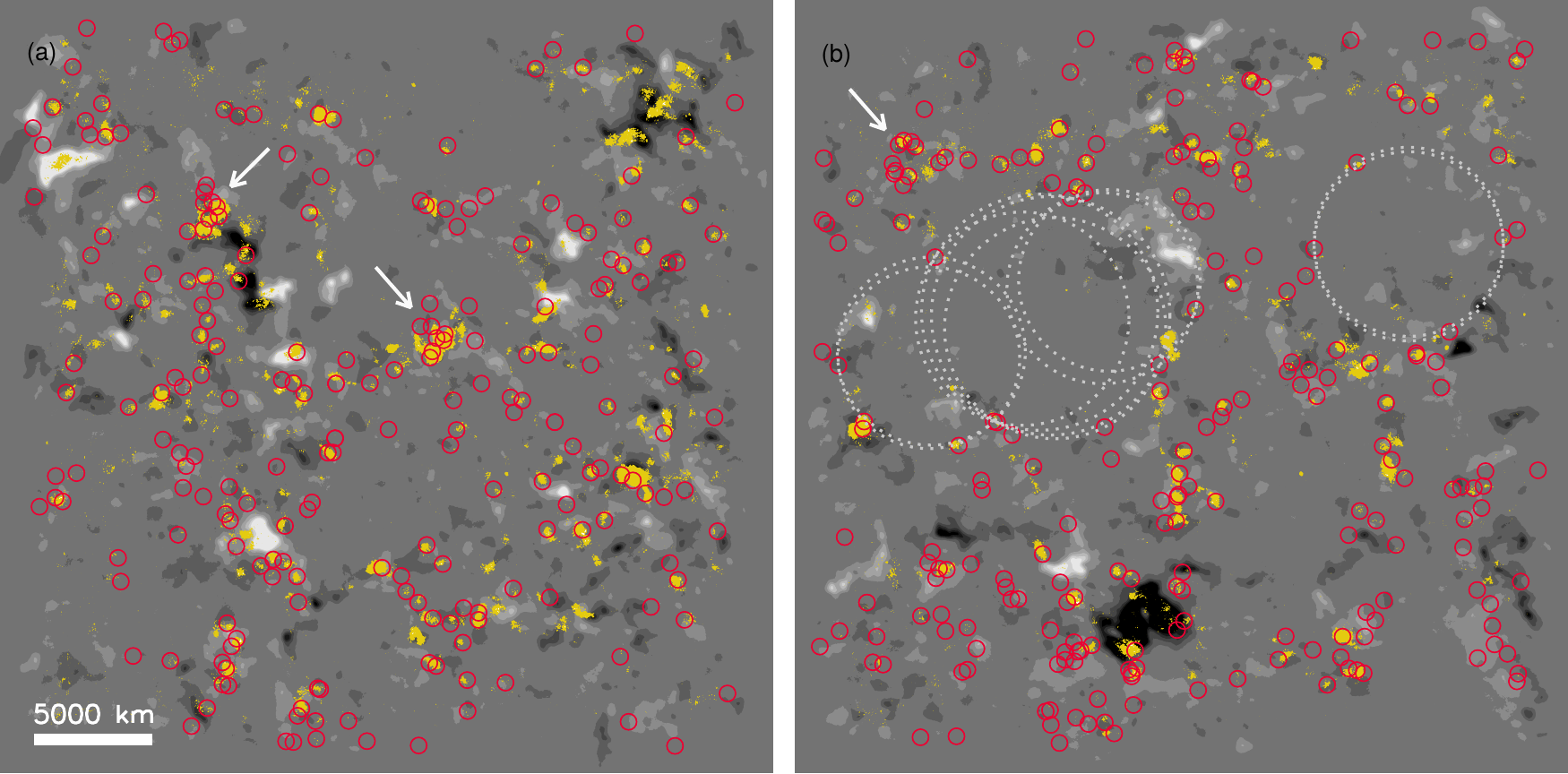}.  In that work,  \cite{martinezgonzalez:12} identify the presence of small-scale magnetic loops in the photosphere as evidenced by opposite circular polarization patches (upward- and downwardly-directed magnetic fields) connected by a detectable linear polarization patch between them (horizontal component of the field).  Such loops were identified in short maps with the \hinode{} SOT/SP (see  \fig{centenoemerge.pdf}), but the high-resolution and high-precision imaging polarimetry of Sunrise/IMaX for the first time permitted this view of their spatial/temporal organization.  This work reveals  persistent areas where none of these small-scale loops appear: ``dead calm'' regions of the quiet Sun.  The authors interpret this result - that the emergence of small-scale magnetic loops in the quiet Sun is not isotropic - as an important observed property of quiet Sun magnetism that must be explained by proposed mechanisms for its generation (such as a small-scale dynamo).

\colfig{deadcalm.pdf}{Two sequences of polarimetric data from Sunrise/IMaX of durations 22 and 31 minutes are presented.  Average  longitudinal magnetograms are shown in greyscale.  During these sequences, small-scale loops in the photosphere (red circles) appear as evidenced  by linear polarization (yellow patches) separating two patches of opposite circular polarization.  Dotted circles (``dead calm'') in the image at right delimit areas where there is very little or no  emergence of these small-scale loops. Arrows point to regions where several loops  appear at the same location in the time series. Figure taken from \citet{martinezgonzalez:12}.}

\subsection{Future Space Missions to Measure Solar Photospheric Magnetic Fields}
\label{sect:future}

\refmark{Perhaps there remain some hidden surprises regarding photospheric magnetic fields that are lurking below our current capability of resolving them because we have certainly not resolved the fine-scale structure of magnetism in the internetwork, and especially in the intergranular lanes.  Both theory and simulations  \cite[e.g.,][]{voegler:07,rempel:14}  and interpretation of observations \cite[e.g.,][]{pietarilagraham:09} suggest that observation of magnetic structure at the dominant scale of photospheric magnetism may be beyond reach with current diagnostic techniques, in recognition of the scattering properties of spectral lines in the photosphere when sampling scales at or below the photon mean free path.} In any event, improvement of adaptive optical systems for the next generation of large ground-based telescopes is proceeding at a rapid pace, so future exploration of photospheric magnetism at very small scales would appear to be most aptly addressed using those facilities.

How then could future space observations contribute uniquely to further the understanding of  photospheric magnetism? Little is still known regarding details of the photospheric field at the extreme solar poles.  The polar regions are important in that it is believed that they play a crucial role in the progression of the large-scale solar dynamo.   Magnetism of the polar regions has been studied extensively using \hinode{} \cite[][and \citeauthor{petrie:15a}, this issue]{ito:10,shiota:12}, but spatial resolution of the extreme poles is limited. Vertical fields are detected mainly by the transverse Zeeman effect and is therefore not very sensitive to weaker fields, and one is detecting fields well above the physical height corresponding to the $\tau_c = 1$ level at disk center because of foreshortening.  The Solar Orbiter mission will fly the Polarimetric and Helioseismic Imager (PHI) in an inclined orbit that will pass within 0.28~AU of the Sun \cite[][this issue]{kleint:15a}.  PHI will perform precision polarimetry of the polar regions near closest approach, so magnetic fields of the polar regions will be attained from a more favorable viewpoint, albeit only for a brief period.

A new frontier for observational studies of solar magnetism is the chromosphere (see \sect{chromag}).  Solar-C is an initiative for a major space mission lead by Japan but, like \hinode{}, having significant contributions from NASA and ESA. The major scientific thrust for Solar-C is the magnetism of the chromosphere, but interpretation of chromospheric field measurements will require accompanying measurements of fields in the photosphere.  The Solar-C concept provides for simultaneous measurements of the field vector using spectral lines forming at several heights from the photosphere through the chromosphere.

In the near term, it is extremely important for the discipline to continue to exploit its excellent resources for photospheric field measurement that are still on orbit. The community should strive to ensure that \sdo{}/HMI and \hinode{} SOT/SP are operational as long as possible in order to build a synoptic data base of vector magnetic fields through a full solar cycle.  One important motivation to sustain these observations is to understand the long-term behavior of non-AR fields: the remnant fields of active regions and also fields apparently not associated with the global dynamo. This is especially important if the most recent extended solar minimum and the present weak maximum heralds a major shift in the dynamo activity, for example the beginning of a new Maunder minimum.

%% file: chromag.tex
\section{Chromospheric Magnetic Field Measurements}
\label{chromag}
\subsection{The Importance of the Chromospheric Field Measurement to the Current Status of Solar Physics }
\label{chromag:importance}

The previous \secs{ground} and \ref{space} demonstrated the high level of sophistication reached especially for the measurement\footnote{In this section, the term measurement refers to the determination of the magnetic field from remote sensing instruments by interpreting signals directly influenced by the solar magnetic field.} of the magnetic field in the photosphere, achieved by using advanced instrumentation in ground-based, balloon-borne and space-based observatories. The spatial and temporal resolution of the measurements in this deepest layer of the solar atmosphere directly accessible by observations has been increased by an order of magnitude within the last three decades. Similarly, recent space observatories operating at wavelengths not accessible from the ground permitted exploitation of the structure of the coronal magnetic field by analyzing the plasma emission of ions trapped in the coronal magnetic field with unprecedented precision. 

The chromosphere is the coupling element between the photosphere and the corona. This extremely dynamic layer is located only one convection-cell size above the solar photosphere ($\approx1000$\,km). Exposed to the cold environment of space, the radiative energy losses effectively cool down the chromosphere. To maintain the typical temperature of approximately 10\,000\,K a substantial amount of energy is required, about 15 times higher \cite[]{aschwanden:07} than the energy necessary to heat the solar corona. Whereas the ``coronal heating problem'' has attracted the attention of the solar physics community for decades, the equally important and interesting problem for the chromosphere has only very recently been emphasized. Similarly to the corona, the candidates for the chromospheric heating are wave dissipation, magnetic reconnection and Ohmic dissipation. Not only are the details of the underlying heating mechanisms poorly understood, but even the relative importance of the possible mechanisms in various structures remains poorly determined. The investigation of the chromospheric magnetism is crucial for both the understanding of the Sun-Earth connection with all its consequences to our technical and natural environment, and the detailed study of fundamental plasma physical processes in density and temperature regimes difficult to achieve in laboratories.

Progress in this field of research has been hindered by several aspects of inferring the properties of the chromosphere. The low gas and plasma densities result in large mean free paths for photons. In this radiatively-dominated regime the population of atomic levels deviates strongly from thermodynamic equilibrium, requiring complex and computationally expensive three-dimensional radiative transfer modeling to interpret the observations. Of similar complexity are magneto-hydrodynamic simulations, which for the photosphere have in the meantime become a reliable and robust tool to understand and analyze the fundamental physical processes. The basis to shed light into this complex field of research are reliable measurements of the chromospheric magnetic field at high spatial and temporal resolution.

\subsection{Challenges to High-Resolution Measurements of Chromospheric Fields}
\label{chromag:challenges}

The chromosphere ``hides'' itself between the observationally easier accessible layers photosphere and corona. With an average height of 500 to 2000\,km above the solar surface it may be observed almost exclusively on the solar disk, with the consequence that the chromospheric information contained in the solar spectrum is to a large extent overwritten by the photospheric emission. The pressure scale height in the solar atmosphere is approximately \revmark{150\,km}, leading to typical densities in the chromosphere of about four orders of magnitude lower than the photospheric ones. This decrease of gas pressure leads to an expansion and subsequently weakening of the magnetic field, rooted in often sub-arcsecond sized magnetic patches in the quiet-Sun photosphere. 

The main tool for diagnosing the chromospheric magnetic field is spectropolarimetry in Fraunhofer lines formed under chromospheric conditions. With a few exceptions, the chromospheric signature in these spectral lines is only present in a narrow range around the line core. Weak fields and the low densities require highly sensitive spectropolarimetric measurements to detect these signals. Additionally, these low densities result in an increase of the typical velocities in the chromosphere: the Alf\'en speed reaches values of 100\kms{}, the sound speed is around 20\kms{}. Gas, plasma and wave motions can easily achieve these velocities, shock waves can form and reconnection of magnetic field lines can occur, making measurements at high temporal and spatial resolution mandatory.

The resulting dilemma was already discussed in \sect{overview} for photospheric conditions. There, the typical velocities are on the order of 5\kms{}, but the velocities in the chromosphere can easily be an order of magnitude higher. As a consequence, the maximum permissible exposure time to freeze in the solar evolution for a large aperture solar telescope ($\gtrsim$1\,m) lies in the \refmark{ten  millisecond range} (see \fig{expt_sn.pdf}a). On the other hand, the photon flux in the line core, i.e., the part of the spectrum containing the chromospheric information, is on the average only 10--20\% of the continuum intensity. Additionally, to characterize the weak chromospheric fields, a high S/N ratio in the range of a few times $10^{-4}$ is desirable. To achieve these goals, highly photon efficient instrumentation on large aperture solar telescopes, not necessarily operated at their diffraction limit, is mandatory. The Daniel K. Inouye Solar Telescope \cite[DKIST,][]{rimmele:08,tritschler:15}, currently under construction on Maui/Hawaii, and the European Solar Telescope \cite[EST,][]{collados:13} are the logical steps in this direction.

Besides these technical challenges, chromospheric conditions introduce a high level of complexity to the interpretation of the polarization signals. Simplifying assumptions, like the local thermodynamic equilibrium (LTE) conditions prevailing in the photosphere, do not hold anymore. Chromospheric absorption and emission processes need to be modeled under non-LTE conditions involving on average 6--12 atomic levels in the most prominent atoms used for chromospheric diagnostics (H, Ca, Mg, and He). The population of the levels in these atoms additionally strongly depends on the incident coronal radiation field. Its anisotropy in the illumination of the chromospheric layer is responsible for creating imbalances in the population of the sub-levels of the atoms, leading to atomic polarization. Long time scales of up to several hours for ionization and recombination processes introduce a memory effect of the chromosphere to past events, complicating the determination of the initial conditions for theoretical models \cite[][this issue]{delacruzrodriguez:15a}.

\subsection{Diagnostic Techniques}
\label{chromag:techniques}

Our knowledge about the chromospheric magnetic field is mainly based on the interpretation of the polarimetric signal in chromospheric spectral lines. The above mentioned difficulties in measuring the polarimetric signal in these lines motivates the search for alternative diagnostic techniques. \refmark{Two possible candidates are:} extrapolations based on photospheric magnetic field maps and measurements in the millimeter and sub-millimeter wavelength range, soon available at high spatial resolution using the ALMA (Atacama Large Millimeter/sub-millimeter Array). 

Extrapolations of the photospheric field rely on the robustness and simplicity of photospheric magnetic field measurements, which can be obtained on a routine basis at higher spatial and temporal resolution than chromospheric measurements. During the past decade a significant improvement on the reliability of these extrapolations has been achieved, for example by including \halpha{} images in the pre-processing of the data \cite[]{wiegelmann:08a}. \refmark{Unfortunately, the principle of extrapolations, based on minimizing the complexity of the magnetic field in the observed atmospheric volume, is not applicable for the most interesting and very dynamic conditions of, e.g., shocks or reconnection sites. To uncover the secrets at these locations, direct measurements are mandatory.}

The linear relation between millimeter wavelength emissions and the chromospheric temperature makes ALMA an ideal thermometer for the solar chromosphere. In addition, in the near future it will perform magnetic field measurements in the chromosphere with a spatial resolution of down to 0.2\arcsec{} based on two physical processes: the Zeeman effect in high-n recombination lines of hydrogen and diatomic molecules, and the influence of the magnetic field to the temperature distribution by suppressing the power of propagating waves. Details about the usage of ALMA for chromospheric diagnostics can be found in this issue \cite[]{judge:15a} and in \cite{wedemeyer:15a}.

In the following sections we return to the current state-of-the-art measurement technique for chromospheric fields using spectropolarimetry in Zeeman and Hanle sensitive spectral lines in the visible and the near infrared regime. Here the improved observational capabilities have led to significant progress during the last couple of years, which will continue in the coming decade with the advent of large-aperture solar telescopes.

\subsubsection{Diagnostic Spectral Lines}

In the strictest sense the term chromosphere is defined as the region of \halpha{} emission. Using the \halpha{} line as the tool to investigate the chromospheric magnetism therefore seems to be an obvious choice. Unfortunately, the above-mentioned difficulties in modeling spectral lines are most prominent in this line. \halpha{} originates from an excited level of the hydrogen atom ($n=3$), whose population strongly depends on the temperature and radiative processes in the atmosphere, and therefore making the interpretation of the spectral profiles exceptionally complex \cite[]{leenaarts:12}.

Spectral lines that are easier to interpret have been identified and are now observed on a routine basis.  \fig{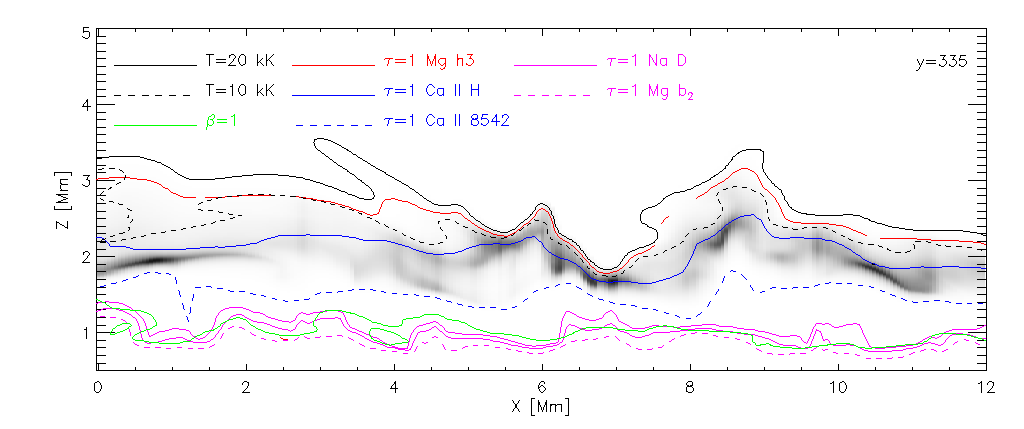} \cite[snapshot from Bifrost chromospheric modeling code, taken from][]{carlsson:15a} presents an overview of these lines which have emerged as especially well suited for diagnosing chromospheric magnetism. Sorted by their formation height from bottom to top, obtained from non-LTE modeling, these lines are: \mgib{} at 517\,nm, the \nad{} line pair at 589\,nm, the \caii{} infrared triplet at 854\,nm, the \caiihk{} lines at 390\,nm, \heid{}  at 588\,nm, the \hei{} triplet at 1083\,nm, and  the \mgiihk{} lines at 279\,nm. The latter has been observed on a routine basis only in spectroscopic and imaging mode by the IRIS mission \cite[]{depontieu:14} and the second Sunrise stratospheric balloon flight \cite[]{barthol:11}. Its potential to retrieve the magnetic field vector is very high, but so far limited to theoretical calculations \cite[]{trujillobueno:14}.

The quiet-Sun modeling underlying \fig{he_10830-058.png} demonstrates the high corrugation of especially the upper chromospheric layers \cite[]{carlsson:15a}. The corrugation becomes even stronger above active regions containing pores or sunspots. The height variation can easily span several Mm and must be taken into account for the correct interpretation of the spectropolarimetric signals.

\colfig{he_10830-058.png}{Formation heights of various chromospheric spectral lines computed from the self-consistent chromospheric modeling code Bifrost \cite[]{gudikson:11}. The snapshot from a time series shows the iso-temperature surfaces (black lines), the plasma-$\beta$=1 layer (green) and the optical depth $\tau$=1 layers for the spectral lines as labeled in the figure. The layer of absorption in the \hei{}~1083\,nm triplet is indicated by the gray-shaded area. Taken from \cite{carlsson:15a}.}

The \caii{} and \hei{} infrared triplets are currently in the focus for the diagnostics of chromospheric magnetism. The analysis of the \caii{} lines has benefited from improvements in the computationally expensive non-LTE modeling \cite[e.g.,][]{leenaarts:09,delacruzrodriguez:12} and the availability of image reconstruction techniques, thereby permitting production of polarimetric maps at the diffraction limit of the modern solar telescopes at high signal-to-noise ratio \cite[]{scharmer:06,cavallini:06,puschmann:12}. Inversions of the radiative transfer equation (RTE) to retrieve the physical conditions in the chromosphere from the measured Stokes spectra are on a good route to become a standard tool for the solar community, similarly to the inversion tools currently available for the photosphere. The progress in this field is summarized in this issue in the article \cite[][this issue]{delacruzrodriguez:15a}.

\subsubsection{The \hei{} 1083\,nm Triplet}

This triplet has gained attraction due to two major improvements:  developments in detector technology allow for spectropolarimetric observations at noise levels down to the $10^{-4}$ range (in units of the continuum), and the theoretical explanation for the linear polarization signal in the absence of magnetic fields \cite[]{trujillobueno:02}. The latter led to the development of easy-to-use inversion codes \cite[]{asensioramos:08a,lagg:09a}. The biggest advantage of the \hei{} 1083\,nm line over other chromospheric lines is the absence of almost any photospheric contamination in the signal. The reason for this lies in the special formation process of this triplet. The \hei{} 1083\,nm line (as well as the \heid{} line) is a transition in the triplet state of the \he{} atom, which is only sparsely populated under normal photospheric and chromospheric conditions. Coronal ultra-violet (UV) radiation with a wavelength shorter than 50.4\,nm (i.e., 24.58\,eV) is required to ionize the \he{} atoms. The subsequent recombination occurs with equal probability either back to the singlet state, or to the triplet state. Since the chromosphere is highly opaque to coronal UV radiation, this ionization process occurs only in the top most layer of the chromosphere, resulting in a thin slab containing \hei{} atoms in the triplet state. To first order, the physical conditions in this thin slab are height-invariant, an assumption validated by the simplicity of the observed Zeeman signals in this triplet. The flip side of this special formation process is the weak absorption signature above quiet Sun regions, where coronal UV illumination is very low.

\subsubsection{\hei{}\,1083\,nm and the Hanle Effect}

A further strength of the \hei{} triplet is the coverage of a wide range of magnetic field strengths: The Land\'e factors of the \hei{} triplet ($g_{\mbox{eff}}=2.0, 1.75, 1.25$) allow for reliable measurements of the magnetic field vector above $\approx$50--100\,G using the Zeeman effect. For higher field strengths the Paschen-Back effect sets in, shifting the Zeeman sub-levels and introducing an asymmetry in the Stokes profiles \cite[]{socasnavarro:04,socasnavarro:05,sasso:06a}. The Hanle effect covers the region of lower field strengths \cite[0.1--100\,G,][]{trujillobueno:02}. Anisotropic illumination of the thin slab by the photospheric radiation field introduces imbalances in the population of the lower and upper levels of the transition, producing a characteristic linear polarization signal. Weak magnetic fields of up to  $\approx$8\,G change the strength and the direction of polarization (Hanle sensitive regime), whereas the linear polarization signals for stronger fields are only sensitive to the field direction (Hanle saturated regime). 

Despite the above-mentioned advantages of the \he{} triplet, the correct interpretation of the signals poses major challenges for the near future. For the correct modeling of the strength of the line, the coronal environment must be taken into account as well as the importance of collisions for the population of the triplet state. The Hanle measurements require an exact determination of the anisotropy of the incident radiation field, influenced not only by the height of the slab, but also by the presence of brightness variations in the photosphere (e.g. sunspots), and the absorption of optically thick \hei{} layers. Additionally, the ambiguities introduced by the Hanle effect must be treated carefully.

The near-infrared spectral region is particularly well suited for the large-aperture ground-based telescopes which were commissioned recently. The reasons are better and more stable seeing \cite[]{turon:70}, less scattered light \cite[]{staveland:70} and less atmospheric extinction. The New Solar Telescope \cite[NST,][]{goode:10} at the Big Bear Solar Observatory with an aperture of 1.6\,m and the GREGOR Telescope on Tenerife \cite[1.5\,m,][]{schmidt:12} are both equipped with instrumentation for magnetic field diagnostics in the near infrared region, with one of the main foci being on imaging polarimetry and spectropolarimetry in the \hei{} 1083\,m triplet.

\subsection{Examples of Current Chromospheric Field Measurement Results}
\label{chromag:examples}

\subsubsection{\refmark{\hei{} 1083\,nm Triplet}}

\refmark{Two examples of recent \hei{} 1083\,nm observations were selected} to demonstrate recent advances in the interpretation of the spectropolarimetric data and the progress in instrumentation, allowing for observations at unprecedented spatial resolution and polarimetric sensitivity in this line. 
\refmark{The first one shows data from the Facility Infrared Spectropolarimeter \cite[FIRS,][]{jaeggli:10} at the National Solar Observatory's Dunn Solar Telescope (DST):} \refmark{\cite{schad:13} presents an observation} of superpenumbral chromospheric fibrils spanning from the penumbra of NOAA AR 11408 to the internetwork regions outside the active region (see \fig{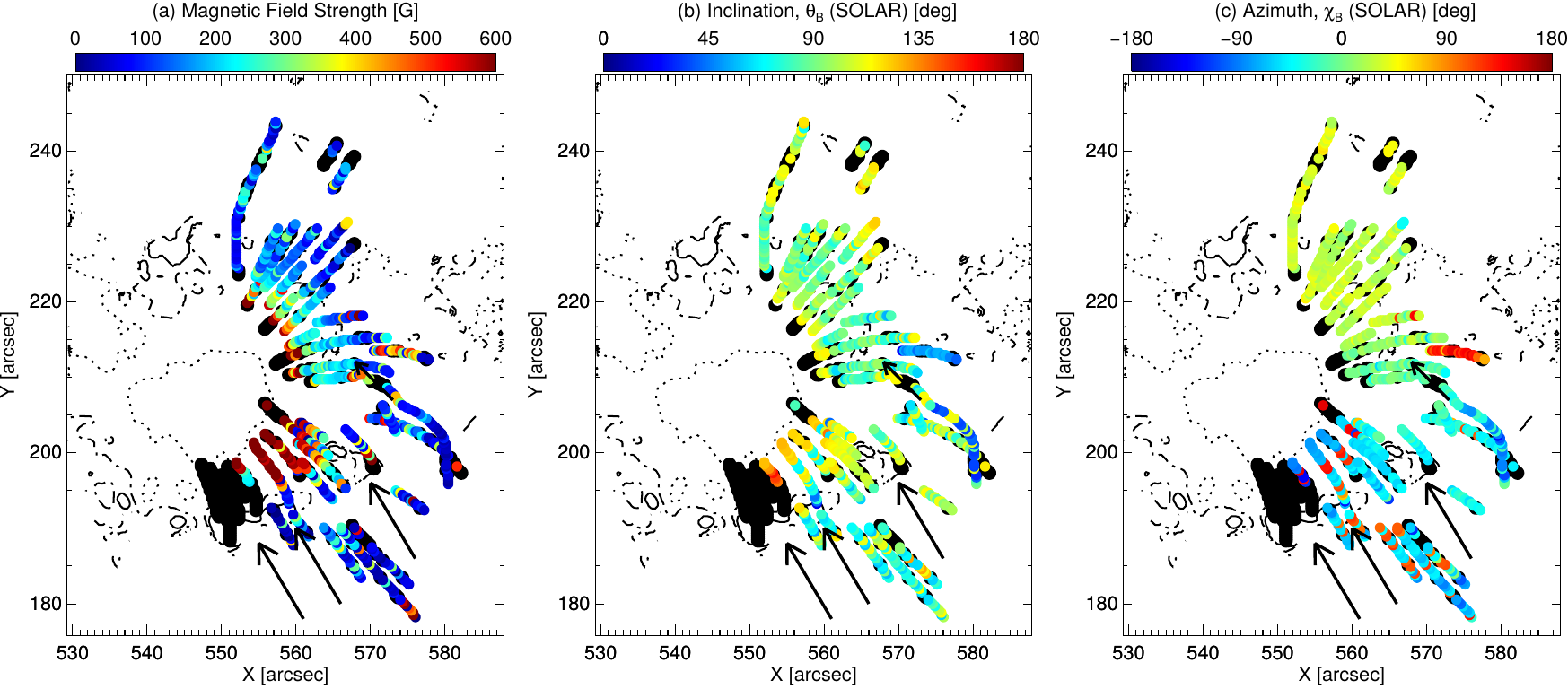}). Limited by the diffraction limit of the instrument of 0.6\arcsec{}, full Stokes maps in \hei{} 1083\,nm were recorded with an average noise level of 3--4\ten{-4} in units of the continuum intensity. The spectra were analyzed with the inversion code HAZEL \cite[``Hanle and Zeeman Light'',][]{asensioramos:08a} which is based on the multiterm calculations of the orthohelium atom of \cite{landi:04}. \citeauthor{schad:13} focussed their analysis on the determination of the magnetic field orientation in these fibrils to answer the question whether the magnetic field vector is aligned with the structures observed simultaneously in \halpha{} and \caii{} infrared intensity maps. This required a thorough analysis to select the correct solution out of up to 8 ambiguous solutions, a consequence of the combination of the well known 180$^\circ$ ambiguity of Zeeman effect with the so called Van Vleck ambiguity \cite[see, e.g.,][]{casini:05a,casini:09a,merenda:06,orozco:15}. 

\colfig{schad_he_fibrils_fig10.pdf}{Magnetic field strength, inclination and azimuth (in solar reference frame) of superpenumbral fibrils observed in the \hei{} 1083\,nm triplet with FIRS at the DST on January 29 2012. The central part of the filament is characterized by horizontal fields below 300\,G, well aligned to the visible fibrils \cite[from][]{schad:13}.}

\refmark{As a result of this analysis, \cite{schad:13} concluded} that the projected direction of the field derived from \hei{} inversions does not deviate by more than $\pm$10$^\circ$ from the visible fibrils. 
For most of the fibrils observed in the \caii{} 854.2\,nm line,  \cite{delacruzrodriguez:11} obtained a similar alignment, but in some of the \caii{} fibrils the magnetic field direction deviates significantly from the visible structures. A thorough investigation of more fibrils at high spatial resolution, ideally simultaneously in \hei{} and \caii{}, is required to resolve this discrepancy.

\colfig{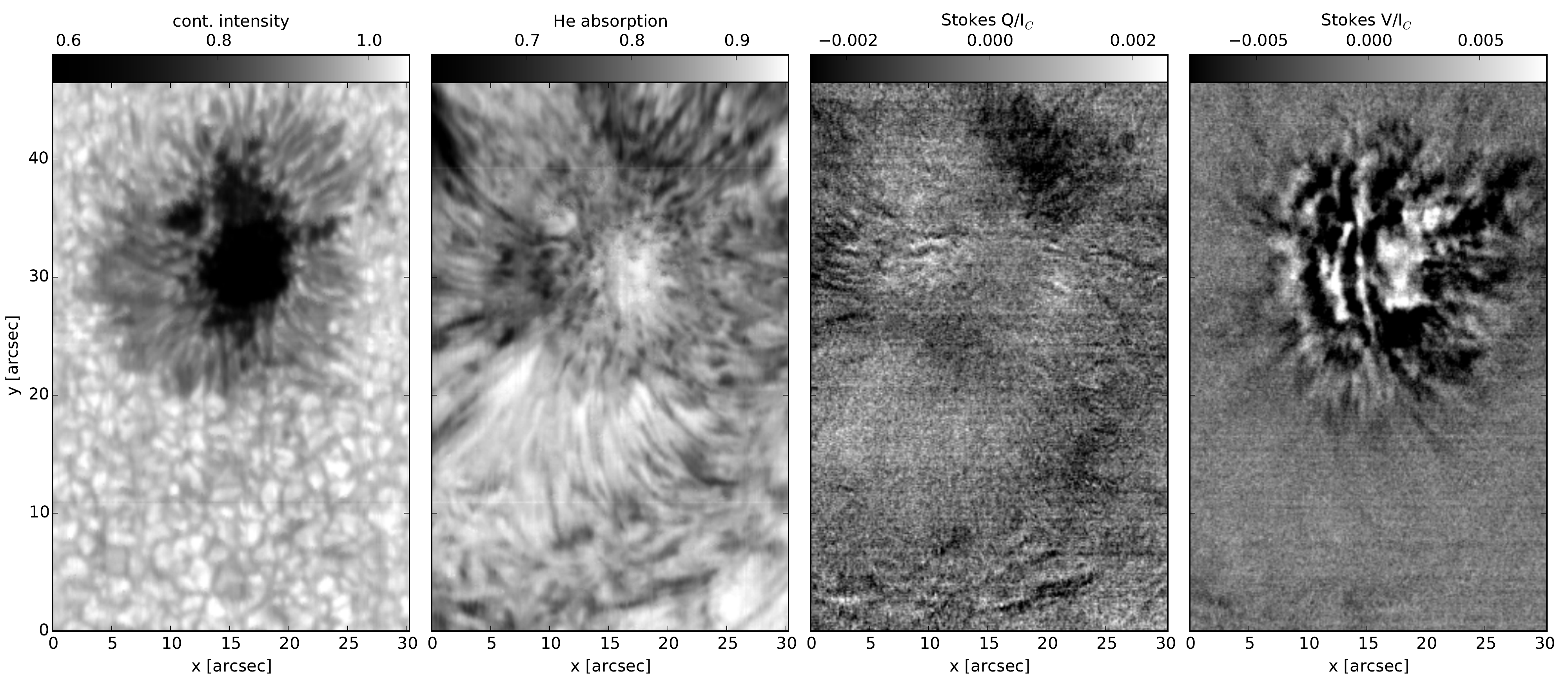}{\refmark{Continuum image and \hei{} 1083\,nm Stokes $I$, $Q$, and $V$ maps of AR~12096 obtained with \revmark{GREGOR/GRIS} on June 27 2014 close to disk center ($\mu=0.98$). The achieved spatial resolution is 0.44\arcsec{} at a spectral resolution of 200\,000. The total exposure time of 1.6\,s per slit position for the full Stokes vector (without readout and processing time) resulted in a noise level of 7--10$\times 10^{-4}$ of the continuum intensity.}}

\refmark{The largest European solar telescope, GREGOR, entered the scientific phase in 2014 \cite[]{schmidt:12}. The GREGOR Infrared Spectrograph \cite[GRIS,][]{collados:12} is producing spectropolarimetric \hei{} 1083\,nm data with an unprecedented spatial resolution close to the diffraction limit of the telescope at signal to noise ratios in the Stokes parameters of up to 3000. An example of such Stokes parameter maps, composed from the scan of the slit over AR~12096 on June 27 2014, is presented in \fig{grisdata.pdf}. A detailed analysis of the fine structure of the chromospheric field is currently being performed at the partner institutions (Kiepenheuer Institut f\"ur Sonnenphysik, Freiburg; Max-Planck-Institut f\"ur Sonnensystemforschung, G\"ottingen; Leibniz-Institut f\"ur Astrophysik, Potsdam; Instituto de Astrof\'isica de Canarias, La Laguna) and will soon appear in a peer-reviewed journal.}

\subsubsection{\refmark{\caii~854.2\,nm}}

\refmark{The highest spatial resolution for spectropolarimetric measurements is currently achievable only using filter-based instruments. As described in \sect{filterinstruments}, these instruments allow to apply image reconstruction routines to fully exploit solar telescopes down to their theoretical resolution limits. The most successful instruments of this type are the CRisp Imaging SpectroPolarimeter at the 1-meter Swedish Solar Telescope \cite[SST/CRISP,][]{scharmer:08} and the Interferometric BIdimensional Spectrometer \cite[IBIS,][]{cavallini:06}.}

\colfig{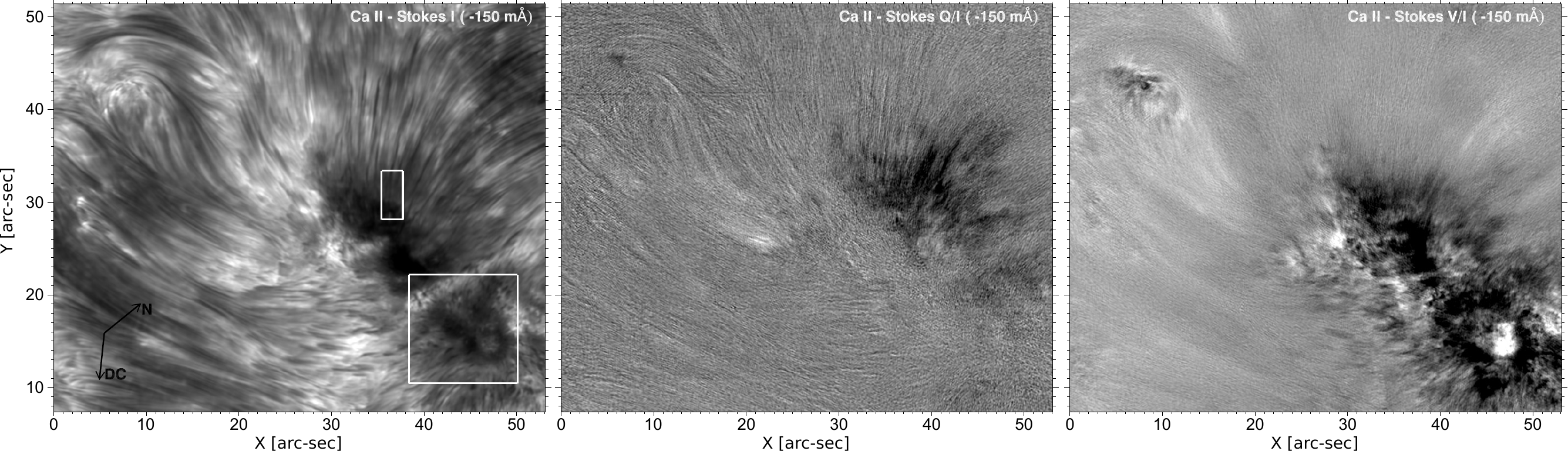}{\refmark{CRISP Stokes $I$, $Q/I$, and $V/I$ filtergrams at $\Delta\lambda=-150$\,m\AA{} from the core of the \caii{} 854.2\,nm line \cite[adapted from ][]{delacruzrodriguez:13}. The boxes mark the regions for which a detailed analysis of umbral flashes and running penumbral waves was performed.}}

\refmark{An example of a chromospheric magnetic field measurement with CRISP is presented in \fig{delacruz_chromo_UF_Fig2.pdf},  showing the Stokes $I$, $Q/I$, and $V/I$ map of AR~11204 observed on May~4 2011. \cite{delacruzrodriguez:13} applied the non-LTE inversion code NICOLE \cite[]{socasnavarro:00} to this dataset and obtained the height dependence of temperature, line-of-sight velocity and magnetic field vector above the umbra and the penumbra of the sunspot at a spatial resolution of $\approx$0.15\arcsec{}. The high temporal cadence of CRISP (16~s for a full-Stokes line scan) enabled the authors to perform a detailed analysis of umbral flashes, revealing their $\approx$1000~K hotter shock front when compared to the surrounding material, without a measurable influence on the magnetic field strength and direction. Additionally, a surprisingly large variation of the magnetic field strength (up to 200~G) in the chromosphere above the penumbra  could be attributed to running penumbral waves. For more examples of high-quality measurements in the \caii{} line using CRISP and IBIS data the reader is referred to \cite[]{pietarila:07} and \cite[]{kleint:12}.}

%% file: challenges.tex
\section{Technology and Data Analysis Challenges}
\label{challenges}

The previous chapters demonstrated that the determination of the magnetic field vector using remote sensing instruments requires a close intermeshing of observational techniques and analysis methods. Technology must be pushed to its limits to obtain data sets with high S/N ratios in all Stokes parameters. An accurate knowledge of all instrumental deficiencies, unavoidably resulting in data degradation, is necessary in order to treat them carefully in the data analysis pipelines. The most common tool for extracting the magnetic field from the observations is the inversion of the Stokes spectra: From an initial guess model atmosphere synthetic Stokes spectra are computed, then the parameters of this model atmosphere are adjusted iteratively until the synthetic Stokes spectra agree with the observed ones \cite[see also][this issue]{delacruzrodriguez:15a}. In this process it is extremely important that the synthetic Stokes spectra are degraded in exactly the same way as the \revmark{instrument alters the spectra originating from the Sun}. Only then the comparison between the observed and synthetic Stokes profile will deliver reliable information about the conditions in the solar atmosphere. 

The aim is, of course, to keep the instrumental deficiencies as small as possible. This is where technological improvements come into play, which have allowed for a significant increase in the accuracy of spectro-polarimetric measurements at high spatial resolution during the last decade. The key ingredients here are advances in detector technology (e.g. FSP, see \sect{groundbased:status}), massively multiplexed focal planes (2-dimensional spectropolarimeters, see \sect{groundbased:obs}), and the use of adaptive optic systems mandatory for operating large aperture solar telescopes at their diffraction limit (see \sect{groundbased:ao}).

In the near future, large-aperture solar telescopes will allow to observe solar features well below scales of 50\,km. Another challenge for the future is to improve the height resolution by performing spectropolarimetric measurements in many spectral lines simultaneously. The difference in the formation height of the lines and in the response to changes of atmospheric parameters will help modern inversion tools to retrieve the stratification in the solar atmosphere with unprecedented accuracy. 

Despite all the efforts to avoid instrumental deficiencies, they will always be present in the data. The proper treatment of these instrumental effects is a big challenge on the data analysis side. \revmark{This has been done successfully during the last years in terms of image reconstruction techniques based on phase-diversity, speckle, or MOMFBD techniques \cite[]{loefdahl:94,vonderluehe:93,vannoort:05}. The next evolutionary steps in this direction} are novel inversion techniques, like the spatially-coupled inversions by \cite{vannoort:12} or the sparse inversion of Stokes profiles by \cite{asensioramos:15}, which take into account the point-spread-function (PSF) of the telescope self-consistently during the inversion process. In a next development step, inversion programs must take into account the natural 2D-coupling between the single-column atmospheres.

But even the cleanest data are subject to ambiguous solutions in the determination of the physical conditions in the solar atmosphere. Such ambiguities are caused by the physical process producing the measured signal (e.g., the well-known 180$^\circ$ ambiguity of the Zeeman effect, or the Van Vleck ambiguity in the Hanle diagnostics), or by the mere presence of a multitude of model atmospheres producing the same Stokes vector within the noise level of the data (degeneracy of the fit parameters).

The resolution of ambiguities is a complex and very often not unique process, involving assumptions about the physical conditions in the solar atmosphere. A successful technique to solve for these ambiguities is to prescribe certain conditions to the solar atmosphere, for example minimizing the divergence of the magnetic field or the vertical current density. A description about available tools for removing the 180$^\circ$ Zeeman ambiguity was given by \cite{metcalf:06}. The solution of the so-called Van Vleck ambiguity, present in Hanle measurements, often requires to choose the ``more likely'' solution, i.e., the solution matching a realistic, physical model of the observed structure \cite[e.g.,][]{orozco:14,schad:13a}.

The identification of the degeneracy in the solution is another important step towards the reliable interpretation of spectropolarimetric data. It helps to identify the simplest model atmosphere compatible with the observed Stokes spectra, which is usually the one with the lowest number of free parameters. Bayesian inversion techniques \cite[]{asensioramos:07} or Markov chain Monte Carlo (MCMC) methods are promising approaches to detect degeneracies.

Ideally, ambiguities and degeneracies in the solutions should be addressed directly by the  observations: multi-line spectropolarimetry can provide the complementary information needed to find the unique solution. Similarly, observations from different vantage points are beneficial. The latter will be achieved for the first time by the out-of-ecliptic mission Solar Orbiter (see \sect{space}), which will provide high-resolution Stokes maps of the photosphere with the Polarimetric and Helioseismic Imager (PHI).

%% file: conclusions.tex
\section{Conclusions}
\label{conclusions}
As mentioned in the introduction, the Sun would be quietly evolving along the main sequence if it were not for its magnetic field. Solar magnetism is responsible for solar activity and space weather events, some of which affect our modern civilization. Magnetic processes appear to be ubiquitous throughout the universe. So for both astrophysical and practical reasons there has been increasing interest in observing and understanding the solar magnetic field. This increase is suggested by \fig{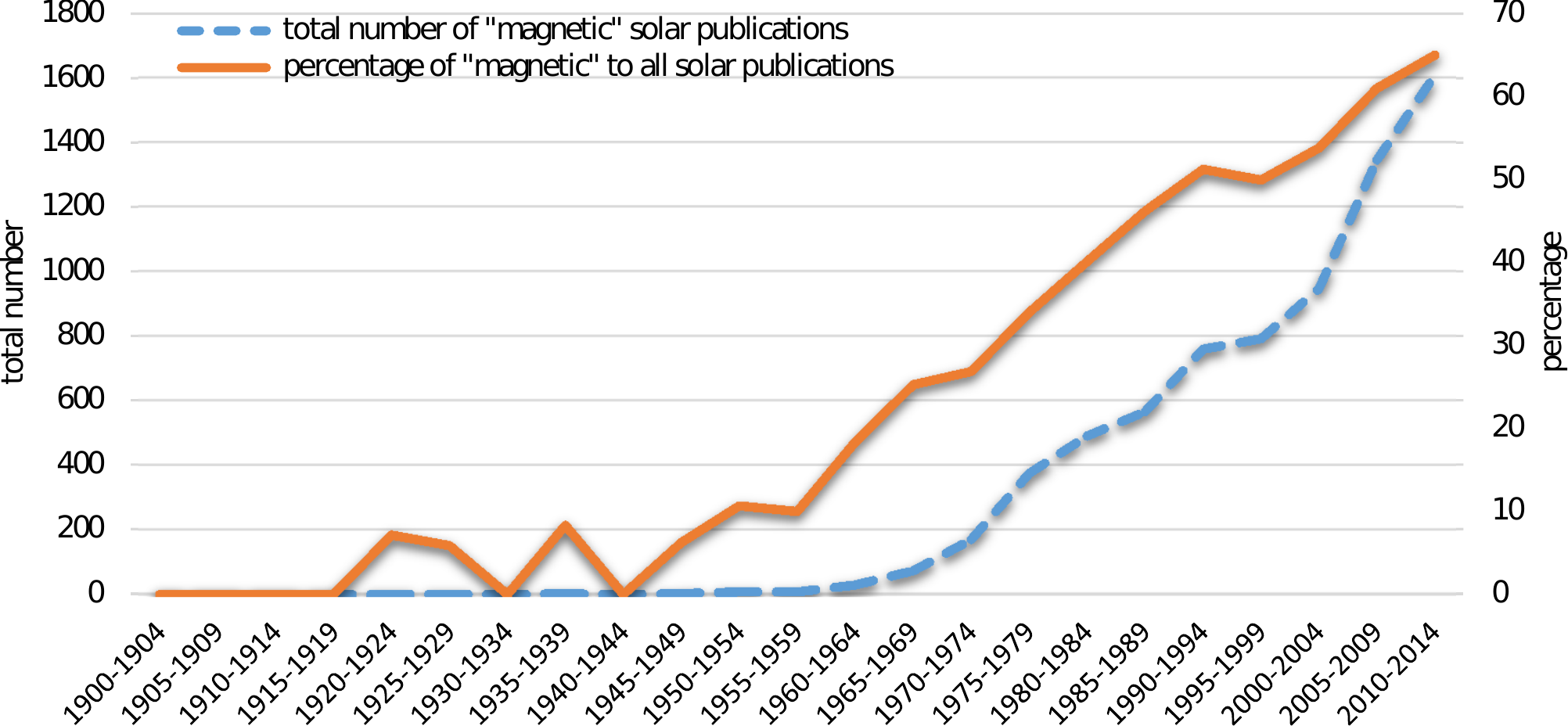} which shows 5-year totals of papers containing ``solar or sun'' and ``photosphere or chromosphere'' and ``magnetic'' in either the title or the abstract of refereed journals (based on NASA ADS). Even more impressive is the increase in the percentage of the publications relative to the ones without the search tag ``magnetic'': during the last 5 years, more than 60\% of all papers on the solar photosphere and chromosphere deal with the subject magnetism.

\colfig{SolMagFields-Pub.pdf}{5-year totals of papers published in refereed journals including the words ``(solar OR sun) AND (photosphere OR chromosphere) AND magnetic''  in the title or abstract according to ADS (dashed blue curve). The solid orange curve indicates the percentage of these publications relative to the ones without the search tag ``magnetic'' (right ordinate axis).}

This impressive growth of interest in solar magnetism stems to a large extent from the recognition of its importance to almost all fundamental physical processes on the Sun and its relevance to space weather, including the efforts to predict space weather events. The rapid development of measurement techniques for solar magnetic fields is a logical consequence, additionally supported by progress in technology, especially in the field of detectors and optical instrumentation. Together with the recent and forthcoming introduction of new telescopes we expect a steady increase of new measurements and understanding of the solar magnetic field. Recent excitement in exoplanet discovery and research has awakened interest in new techniques of spectropolarimetry of stars that should benefit solar observational methods. In the entire two-century history of solar polarimetry, the next decade should be the most exciting and scientifically productive.

%% file: acknowledge.tex
\begin{acknowledgements}
\hinode{} is a Japanese mission developed and launched by ISAS/JAXA, collaborating with NAOJ as a domestic partner, NASA and STFC (UK) as international partners. Scientific operation of the \hinode{} mission is conducted by the \hinode{} science team organized at ISAS/JAXA. This team mainly consists of scientists from institutes in the partner countries. Support for the post-launch operation is provided by JAXA and NAOJ (Japan), STFC (UK), NASA, ESA, and NSC (Norway).
B. Lites was supported in part by the FPP project at LMSAL and HAO under NASA contract NNM07AA01C. The National Center for Atmospheric Research is sponsored by the National Science Foundation.
\refmark{The National Solar Observatory is operated by the Association of Universities for Research in Astronomy (AURA, Inc.) under a cooperative agreement with the National Science Foundation.}
\refmark{The 1.5-meter GREGOR solar telescope was built by a German consortium under the leadership of the Kiepenheuer-Institut f\"ur Sonnenphysik in Freiburg with the Leibniz-Institut f\"ur Astrophysik Potsdam, the Institut f\"ur Astrophysik G\"ottingen, and the Max-Planck-Institut f\"ur Sonnensystemforschung in G\"ottingen as partners, and with contributions by the Instituto de Astrof\`isica de Canarias and the Astronomical Institute of the Academy of Sciences of the Czech Republic.}
\refmark{This study is supported by the European Commission's FP7 Capacities Programme under the Grant Agreement number 312495.}

\end{acknowledgements}